# TRANSPORT OF COSMIC-RAY PROTONS IN INTERMITTENT HELIOSPHERIC TURBULENCE: MODEL AND SIMULATIONS

Fathallah Alouani-Bibi[1] and Jakobus A. le Roux[1,2]

[1]Center for Space Plasma and Aeronomic Research, University of Alabama in Huntsville, Huntsville, USA, fb0006@uah.edu

[2]Department of Space Science, University of Alabama in Huntsville, Huntsville, USA

## Abstract

The transport of charged energetic particles in the presence of strong intermittent heliospheric turbulence is computationally analyzed based on known properties of the interplanetary magnetic field and solar wind plasma at 1 Astronomical Unit (AU). The turbulence is assumed to be static, composite, and quasi-three-dimensional with a varying energy distribution between a one-dimensional Alfvénic (slab) and a structured two-dimensional component. The spatial fluctuations of the turbulent magnetic field are modeled either as homogeneous with a Gaussian probability distribution function (PDF), or as intermittent on large and small scales with a $q$-Gaussian PDF. Simulations showed that energetic particle diffusion coefficients both parallel and perpendicular to the background magnetic field are significantly affected by intermittency in the turbulence. This effect is especially strong for parallel transport where for large-scale intermittency results show an extended phase of subdiffusive parallel transport during which cross-field transport diffusion dominates. The effects of intermittency are found to depend on particle rigidity and the fraction of slab energy in the turbulence, yielding a perpendicular to parallel mean free path ratio close to 1 for large-scale intermittency. Investigation of higher order transport moments (kurtosis) indicates that non-Gaussian statistical properties of the intermittent turbulent magnetic field are present in the parallel transport, especially for low rigidity particles at all times.

## 1. Introduction

Upon entering the heliosphere, galactic cosmic-rays, especially those with a gyroradius of the order or less than the heliospheric scale size of ~100 AU, experience the effect of the interplanetary magnetic field (IMF) (Parker 1958). The IMF, with field lines extending from the solar surface, is continuously advected radially away from the sun at the local solar wind speed. This outward motion combined with the solar rotation is responsible for the IMF's spiral shaped morphology (Parker 1958, Wilcox & Ness 1965). The properties of the solar wind and IMF, such as the speed and intensity respectively, change over the solar cycle. The solar wind speed shows an additional helio-latitudinal dependence during solar minimum activity periods. These changes lead to time variations in the modulation of the cosmic-ray intensity partly because cosmic-ray transport is affected by changing scattering conditions during interaction with magnetic field fluctuations throughout the heliosphere (e.g. Jokipii, 1966, Fisk, 1971, Mavromichalaki et al. 1988, McDonald et al. 1997, le Roux & Potgieter, 1995, Sabbah 2000).

Observations (Marsch & Liu 1993, Burlaga et al. 1993, Marsch & Tu 1994, Forsyth et al. 1996, Horbury et al. 1997, Burlaga 2001, Burlaga & Viñas 2004, Bruno et al. 2004, 2007, Burlaga et al. 2007, D'Amicis et al. 2010, Wan et al. 2012, Alouani-Bibi & le Roux 2013) have shown that the interplanetary magnetic field is highly intermittent with multi-scale structures whose statistical properties are non-Gaussian (Tsallis 1988, 2004). The time scales over which the magnetic field turbulence is non-Gaussian ($q$-Gaussian), extends from the shortest time scales of ~1 min closer to the turbulence dissipation range to time scales ~100 days far exceeding the decorrelation time of the turbulence. These aspects of the solar wind magnetic field statistics can have important implications for how energetic particles propagate in the solar wind. Currently, only the first two moments in particle displacement are typically considered when modeling energetic particle transport in idealized uniform turbulence. However, higher moments might also be needed to properly capture the characteristics of particle transport in more realistic intermittent turbulence as is discussed below.

Particles entering intermittent patches of strong magnetic field fluctuations are found to have a slower diffusion rate. These particles are trapped in coherent magnetic field structures for a longer time due to strong pitch-angle scattering. This can result in a significant departure from classical Gaussian transport in terms of subdiffusion or compound diffusion during intermediate time periods of considerable duration. Particle transport reflects the properties of the intermittent magnetic field. For such non-Gaussian transport, a classical Fokker-Planck description, where pitch angle scattering consists of continuous small-angle variations, may no longer be valid (Pommois et al. 2007).

Intermittent turbulence can strongly change the efficiency of particle transport compared to uniform turbulence, especially for parallel transport as is discussed below in Section 3.2. This needs to be taken into account when inferring some of the properties of cosmic rays (CRs), e.g.

residency times either within the heliosphere or in our galaxy, the ratio of primary to secondary CR populations, and the modulation of cosmic rays in the heliosphere. Additionally, particle shock acceleration, especially at nearly perpendicular shocks (Zank et al. 2001), can be affected by the variation in turbulence geometry. This can change from transverse incompressible composite turbulence upstream to isotropic and compressive downstream. Moreover, the effect of non-Gaussian intermittent turbulence upstream and downstream (Burlaga & Ness 2009) on shock acceleration still needs to be explored. These aspects of turbulence are intimately linked to the dynamics of the shock which in turn determines the efficiency of shock acceleration.

Solar wind turbulence is generally described using one of the following two approaches: (1) a composite quasi-3D model supported by theory and observations (Matthaeus et al. 1990, Zank and Matthaeus, 1992, 1993, Bieber et al. 1994, 1996) whereby turbulence to lowest order has a minor one-dimensional (1D) Alfvénic component (referred to as "slab") with the wave vector $\vec{k} \parallel \vec{B}_0$, $\vec{B}_0$ being the background magnetic field, and a dominant two-dimensional (2D) structured component for which the wave vector $\vec{k} \perp \vec{B}_0$. Both the slab and 2D components have magnetic fields orthogonal to $B_0$. The anisotropy in the second model depends on the distribution of energy between the parallel (slab) and the perpendicular (2D) modes. Typical energy partition between the slab and 2D component is generally of the order of 20% for slab and 80% for 2D of the total turbulence energy (Bieber et al. 1994, 1996). (2) The critical balance model ($\delta v k_\perp / v_A k_\parallel \sim$ 1, where $\delta v$ is the root-mean-square of the plasma velocity fluctuations, $v_A$ is the Alfvén speed, $k_\parallel$ and $k_\perp$ are the turbulence wavenumbers parallel and perpendicular to the mean magnetic field) (Goldreich & Sridhar 1995) where the parallel wave mode cascade is weakened relative to the cascade of the perpendicular mode based on the assumption that the nonlinear time is equal to the Alfvén crossing time (strong turbulence). This turbulence model, which has some support from MHD simulations (e.g., Lazarian et al., 2012, Beresnyak 2012), predicts a Kolmogorov slope for the power spectral density (PSD) in the perpendicular direction and an anisotropy relation $k_{\parallel} \propto k_{\perp}^{2/3}$. It has been argued to have important consequences for cosmic-ray transport (e.g., Beresnyak et al. 2011, Lazarian et al. 2012). Recently, asymmetric superdiffusion of magnetic field lines in anisotropic MHD turbulence connected to Richardson diffusion has been discussed (Beresnyak 2013) whose consequences for cosmic-ray transport need to be investigated in future work. However, note in a more recent observational study that Forman et al. (2011) used Ulysses magnetic field data to investigate both the composite and critical balance models of solar wind turbulence. The authors found that both models gave a good reproduction of the observed scaling of magnetic wave power with angle and frequency. It was concluded that the composite turbulence model, which has the advantage of mathematical simplicity, cannot be ruled out. Thus, in this paper we adopt for simplicity the composite (slab + 2D) formalism as the framework for our computational model to take a first step in studying the effect of intermittent magnetic field turbulence on cosmic-ray transport, which is a complicated problem. The more advanced critical balance turbulence model will be explored in future work.

The structure of this paper is as follows: In Section 2 we present our numerical model, the algorithm for the generation of turbulence and discuss the different geometries and scales used for the non-Gaussian intermittency. In Section 3, we present the results and discussion of our simulations for uniform and intermittent turbulence, and comparisons with existing theories. This followed by the conclusions in Section 4.

## 2. Numerical model

The scattering and transport of particles in phase space $(\vec{x}, \vec{p}, t)$ is tracked by solving the Newton-Lorentz equation of motion:

$$\frac{d\vec{U}}{dt} = \vec{U} \times \vec{\Omega}. \tag{1}$$

where $\vec{U} = \gamma\vec{v}$ is the relativistic particle proper velocity and $\vec{\Omega} = e\vec{B}/\gamma mc$ is the particle gyro-frequency. The transport of particles is considered within the test-particle limit where the back-reaction of energetic charged particles on the plasma is assumed to be negligible due to their relatively low number density. Given the rigidity range considered in our simulations and an Alfvén speed $v_A$ of ~40-50 km/s in the supersonic solar wind, we are in the fast particle limit $v/v_A \gg 1$ and we can assume to good approximation that the effect of the motional electric field on energetic particle transport is negligible when compared to the magnetic field, and that energetic particles experience the magnetic field as static. Furthermore, the magnetic field is assumed to be a combination of a uniform background magnetic field with a mean value of $B_0 = 5$ nT which is appropriate for 1 AU, and a turbulent magnetic field component transverse to the mean magnetic field with a zero mean given by

$$\vec{B}(x, y, z, q) = B_0 \hat{z} + \vec{\delta b}(x, y, z, q) \tag{2}$$

The assumption of a transverse turbulent field component implies that the field is incompressible and ensures that the field fulfils the required divergence-free condition. An incompressible field is a justified approximation for quiet solar wind conditions where observations showed density fluctuations to be small (Matthaeus et al. 1991). Compressive turbulence was found to be important on short time scales downstream of shocks, such is the case in the inner-heliosheath (Burlaga & Ness 2009), but is beyond the scope of this paper.

In addition to the spatial coordinates, the turbulent field is also assumed to depend on the entropic index $q$ to account for the non-Gaussian property of the field PDF at smaller scales. Following the notation used in Giacalone & Jokipii (1994, 1999), the turbulent field is expressed as a sum over a series of linear plane waves with random phase $\beta_n$:

$$\vec{\delta b}(x,y,z,q) = \chi(q,z)\,\text{Re}\left[\sum_{n=1}^{N}\vec{\delta b}(k_n)\,e^{i(k_n z'(x,y,z)+\beta_n)}\vec{\xi}_n(\hat{x},\hat{y})\right] \quad (3)$$

This formulation is equivalent to a shifted discreet Fourier transform in the limit of a vanishing wave power spectrum beyond the assumed higher and lower $k$-value cut-offs. In Equation (3), $k_n$ is the wavenumber of the $n^{th}$ the turbulence mode. The direction of the turbulent field is defined as

$$\vec{\xi}_n(\hat{x},\hat{y}) = \cos(\alpha_n)\hat{x}' + i\sin(\alpha_n)\hat{y}' \quad (4)$$

where $\alpha_n$ is the polarity angle. The quantities with superscript (') are readily transformed to the simulation frame by applying a double rotation using standard Euler rotation matrices, i.e.:

$$\vec{V}' = R_y(-\theta_n)R_z(-\phi_n)\vec{V}' \quad (5)$$

For the case of slab turbulence, we assume $\theta_n = 0$, while the rest of the phase angles ($\alpha_n^{sb}$, $\beta_n^{sb}$, $\phi_n^{sb}$) are generated randomly in the open interval $]0, 2\pi[$. For 2D turbulence both $\theta_n$ and $\alpha_n$ are set to $\pi/2$, while ($\beta^{2D}$, $\phi_n^{2D}$) are generated randomly in $]0, 2\pi[$ (Giacalone & Jokipii 1994, 1999). We define a magnetic field realization as being a set of $N$ randomly generated phase angles, i.e. ($\alpha_n^{sb}$, $\beta_n^{sb}$, $\phi_n^{sb}$) for slab and ($\beta_n^{2D}$, $\phi_n^{2D}$) for 2D turbulence ($N$ is the maximum number of wave modes used in Equation (3)). Knowledge of this set fully defines the turbulent field orientation at any position in space. Based on simple dimensional analysis the wave-vector spectrum is chosen as:

$$G(k) = \frac{A k^s}{1 + (kL)^{\gamma+s}} \tag{6}$$

where $\gamma = 5/3$ is the Kolmogorov slope index, $s$ is the energy-containing range power-law index, $L$ is the bend-over scale length, and $A$ is a normalization constant. We follow the models by Matthaeus et al. (1990), Zank and Matthaeus (1992, 1993), and Bieber et al. (1994, 1996) whereby low-frequency solar wind turbulence during quiet times can be modeled to lowest order to be composed of a 2D component representing coherent or random magnetic field structures (Figures 1(a), 1(b), and 1(c)) coexisting with an Alfvénic 1D slab component (Figure 1(d)). The bend-over scales are taken to be $L_{c,sb} = 0.03$ AU for slab turbulence and $L_{c,2D} = 0.1\ L_{c,sb}$ for the 2D component. We assumed in our simulations a flat energy range for slab turbulence ($s = 0$) and a decaying energy range for the 2D component ($s = 1$) (Giacalone & Jokipii 1999, Matthaeus et al. 2007).

The function $\chi$ in Equation (3) accounts for the effect of intermittency and non-Gaussian statistics on the magnitude of the turbulent field. This function is derived to be scale dependent in accordance with observations (Alouani-Bibi & le Roux 2013). A study of intermittent turbulence assuming a Gaussian PDF has been reported in Laitinen et al. (2012). Observations, however, show that departure from Gaussian statistics is a key signature of intermittency in the solar wind. Therefore the function $\chi$ itself is assumed to have a $q$-Gaussian character, i.e. it has a probability distribution function for a given entropic-index $q$ given by:

$$PDF(\chi) = \frac{\sqrt{\beta}}{C_q} \left(1 + (q-1)\beta \chi^2\right)^{-1/(q-1)} \tag{7}$$

where $\beta = 1$ and $C_q = \sqrt{\pi/(q-1)}\,\Gamma((3-q)/(2q-2))/\Gamma(1/(q-1))$. In the present model only intermittency in the magnitude of the turbulent magnetic field is considered. A more general approach could include intermittency in both the direction and the magnitude of the field, which means a non-Gaussian variation of the polarity vector $\xi_n$. Nevertheless, directional intermittency is partially included in our model because the direction of the local magnetic field is defined by the combination of the z-axis-aligned background field and the intermittent $q$-Gaussian turbulence. For strong turbulence, as is the case for the present study, particles will see variations in the direction of the local field that are themselves $q$-Gaussian. The corresponding $q$-Gaussian distributed amplitudes can be generated using either the Box-Muller method (Thistleton et al. 2007) or by inverting the function in Equation (7). Both approaches yield similar results, particularly in the tail of the PDF. We chose the range of variation of the turbulent field from $\delta b_{min} = 0.1 B_0$ to $\delta b_{max} = 10 B_0$ when generating intermittency using the latter inversion approach so that

$$\chi(q,z) = \left[\frac{1}{(q-1)\beta}\left(\frac{C}{\sqrt{\beta}}(1-r(z))^{-(q-1)} - 1\right)\right]^{-1/2} \quad (8)$$

where $r(z)$ is a uniformly distributed random number in the interval $[r_{min}, r_{max}]$ according to

$$r_{min,max} = 1 - \frac{\sqrt{\beta}}{C_q}\left(1 + (q-1)\beta\,\delta b_{min,max}^2\right)^{-1/q-1} \quad (9)$$

The value of $r$ is assumed constant, while the coordinate $z$ determines the position within a given intermittency patch along the $z$-axis. The related values of the entropic index $q$ and the intermittency patch size $L_P$ are both derived from observations (Alouani-Bibi & le Roux 2013). For a given patch size $L_P$, the intermittency is constructed by assigning to each of the $N$ patches along the z-axis, where $N \sim \lambda_{max}/L_P$, a distinct value of $\chi$ (Equation 8). $\lambda_{max}$ is the longest turbulence mode wavelength assumed in our simulation (see below). Two intermittency time scales are considered based on ACE observations (Alouani-Bibi & le Roux 2013). These are τ ~ 0.2 hr and 9 hr. The corresponding non-Gaussian entropic indices for these time scales are $q$ = 1.55 and $q$ = 1.27 valid for the solar maximum epoch 2001-2002. These time scales are chosen short enough to be well inside the maximum time scale for which the magnetic field PDF is non-Gaussian. Based on both the chosen time scales and the mean flow speed during solar maximum, we get two intermittency scale sizes of $L_P = 0.1 L_{c,sb}$ and $3.3 L_{c,sb}$, respectively. The entropic index used in our simulations is therefore scale dependent. This is an improvement on previous results by Alouani-Bibi & le Roux (2012) where the values for $q$, and $L_P$ were independent and chosen arbitrarily. In the context of MHD turbulence, other forms of turbulence intermittency were analyzed, such as polarization intermittency (Beresnyak & Lazarian 2006, Beresnyak 2012, Lazarian 2012) which describes the scale-dependent changes in the direction of interacting Alfvénic modes.

Energetic particle transport coefficients were calculated using different numerical schemes for the integration of the particle trajectory, e.g. higher order Runge-Kutta (RK) methods and the Bulirsch-Stoer (BS) method both with step size and error control (Engeln-Müllges & Uhlig 1996, Press 1992) for a self-consistency check. The results presented in the next section were obtained using the Bulirsch-Stoer method with a Richardson rational extrapolation (Press 1992). In terms of computational efficiency and accuracy, methods such as Bulirsch-Stoer or Prince-Dormand RK8(7) are marginally different. For lower order methods that lack adaptive time-stepping and error control this becomes an issue. Comparing simulations using the BS method, and the RK4 method with a fixed time step showed large differences in efficiency and accuracy. For the same initial parameters, the relative error in the particle energy was ~$10^{-4}$ for RK4 while for BS it was ~$10^{-12}$. In addition, both methods reached the same physical time after 4.5 hours of computational time for the BS method and 9.2 hours of computational time for the RK4 method. It is important to note that for models such as (BS) and RK8(7) the maximum error is set by the tolerance chosen at the start of the simulation, which is fulfilled at each time step. For RK4, on the other hand, the error tends to grow as the simulation time increases, making the performance of RK4 worse than either (BS) or RK8(7) over longer time periods. Unfortunately, one needs an extended simulation time in order to reach a diffusion state for the transport coefficients. An alternative would be to choose small enough initial time step for RK4 to have a reasonable error in particle energy over the diffusion time. Unfortunately, this will tend to slow down the simulations substantially.

## 3. Results and discussion

For the numerical results shown in this section, 2240 particle trajectories were simulated in each of the 32 independent magnetic field realizations. The turbulent field is calculated at the particle position at each time step as described in the previous section. To ensure that the longest wavelength turbulence modes fit inside the simulation domain, the simulation box size is set equal to the longest wavelength of the turbulence used in our simulation (Tautz 2010). Thus the minimum turbulence wavelength is $\lambda_{min} = 10^{-3}L_{c,sb}$ and the maximum wavelength is $\lambda_{max} = 2 \times 10^{4}L_{c,sb}$, while a total of 512 logarithmically spaced turbulence wave modes are calculated to ensure a well resolved turbulence spectrum. Although the maximum wavelength $\lambda_{max} = 600$ AU so that it appears not to fit inside the heliosphere, note that this is the wavelength along the mean magnetic field. Based on the expression for the path length along the interplanetary spiral magnetic field given by

$$L = -\frac{U_{sw}}{2\Omega}\left[\Phi\sqrt{1+\Phi^2} + \log\left(\Phi + \sqrt{1+\Phi^2}\right)\right] \quad (10)$$

with $\Phi = -\Omega(r-r_0)/U_{sw}$, the path length can be estimated. Assuming the generic values of $U_{sw} = 450$ km/s for the solar wind flow speed, $\Omega = 2.7\times10^{-6}$ rad/s for the solar angular speed, and $r_0 = 0.047$ AU for the source distance of the field (assumed to be at 10 solar radii from the Sun), a path length $L$ of 600 AU can be reached at a radial distance $r \sim 36.5$ AU from the Sun, which is well within the termination shock boundary.

We considered particles with a normalized gyroradius (normalized rigidity) $R = R_g/L_{c,sb}$ at $90°$ pitch angle varying from 0.025 to 0.25. Five values were considered inside this interval, i.e. $R$ = [0.025, 0.05, 0.075, 0.1, 0.25]. The assumed box size ($\lambda_{max}$) puts a constraint on the computational time for which a particle of certain energy can be tracked. Particle transport coefficients are compared up to the same physical time, which is the shortest time it takes for any particle to leave the simulation box. This time can be estimated as the time it takes for a free streaming particle with the highest rigidity ($R$ = 0.25) to cross a length $\lambda_{max}$ which is $t_{FS} = 4 \times 10^4$ $\Omega^{-1}$. Because of the strong turbulence considered here ($\delta B/B_0 \sim 1$) such a free streaming scenario is unlikely. We found, by looking at the farthest point in the trajectory of each particle for all field realizations, that no particle has left the simulation domain within a time of $t_{max} = 1.25\times10^5$ $\Omega^{-1}$. This is particularly fortunate as the time $t_{FS}$ would not have allowed enough time for particles transport to converge to a steady state transport regime along the background field.

A sample of the magnetic field generated by our model is shown in Figure 1, representing the magnetic field lines and the field vectors for the composite turbulence model (Figures 1(a), 1(b) and 1(c)) and the field lines for the slab model (Figure 1d). The impact of intermittency is shown on the magnetic field lines for small-scale intermittency (see the central panels in Figures 1(a), 1(b) and 1(d)), and for large-scale intermittency (right panels of Figure 1(a), 1(b) and 1(c)). 2D magnetic field structures can be seen in Figures 1b and 1c for the case of composite turbulence in both the projected field lines and the field vectors. The intermittency of magnetic field turbulence is assumed only along the $z$-axis. Thus, for a given $z$-coordinate the 2D component is Gaussian along the perpendicular $xy$-plane. To include intermittency in the $xy$-plane will lead to sharper boundaries between the 2D structures shown in Figure 1(c) (e.g., Greco et al. 2009), but such an analysis is beyond the scope of this paper. Figure 1 represents the magnetic field generated using a single field realization. Figure 1(a) suggests that particle diffusion along the mean magnetic field ($z$-direction) will be slowed down in the case of intermittent turbulence because of more extensive large-scale field-line wandering transverse to the background magnetic field but also because of stronger pitch-angle scattering by the enhanced amplitude smaller-scale field fluctuations. This interpretation is supported by our particle simulations discussed below. Figure 1(b) does not provide such a clear cut insight on whether perpendicular diffusion of energetic particles will be enhanced or reduced in intermittent turbulence when compared to uniform turbulence. The field lines in the intermittent turbulence cases are more spread out in the plane transverse to the mean magnetic field, suggesting that the enhancement in field-line wandering will be benefit perpendicular transport. On the other hand, the strong presence of large closed field-line structures in the case of intermittent turbulence that trap particles can reduce perpendicular transport. The simulation results discussed below reflect this competition since both an enhancement and a reduction in perpendicular transport were found.

### 3.1. Uniform turbulence

Figure 2(a) shows the parallel diffusion coefficient as a function of time for particle rigidities $R = 0.025$ and $R = 0.25$. The parallel diffusion coefficient is defined as $\kappa_\| = <(\Delta z)^2>/(2t)$, where the ensemble averaging is carried out for all particles of a given rigidity and for all field realizations. The figure also compares the slab and composite (2D plus slab) turbulence models. Two configurations for the composite model are chosen, each representing a different ratio of the 2D to slab energy, ($\delta B^2_{2D}/\delta B^2_{sb} = 1$ and 4), but the total energy in the turbulence is kept constant at the level of strong turbulence in each case so that $(\delta B^2_{2D}+\delta B^2_{sb})/B^2_0 \sim 1$. This level of turbulence is used in all the results discussed in this paper, and represents an upper limit for the turbulence level in the solar wind magnetic field at 1 AU.

The behavior of the diffusion coefficient for the two models is consistent with previous results (Giacalone & Jokipii 1999, Qin et al. 2002a, 2002b, Tauz et al. 2006, Alouani-Bibi & le Roux 2012). That is, particles first undergo a superdiffusive phase lasting about $10^2\Omega^{-1}$ where particles did not have enough time to sample the structure of the turbulence. This is followed at late times by a diffusive regime ($\kappa_\| \sim$ const. or $<(\Delta z)^2> \sim t$). The value of $\kappa_\|$ at late times is approximately inversely proportional to the energy in the slab component for the high rigidity case, so that for pure slab turbulence the value of $\kappa_\|$ is at its lowest. Thus the results are close to being consistent with quasi-linear theory (QLT) for particle propagation in slab turbulence (see Eq. (11)) suggesting that 2D turbulence has little effect on parallel particle transport. This is confirmed by QLT for particle transport in static uniform axisymmetric 2D turbulence which predicts that the 2D component has a negligible effect on particle scattering (e.g., Bieber et al. 1994, le Roux et al., 2004). However, the difference between the diffusion coefficient for pure slab and composite turbulence models is less pronounced in the low rigidity case. This suggests that increasing the 2D component counteracts the strong increase in diffusion coefficient due to the reduction in the slab wave power at low rigidities. For such particles the gyroradius is smaller than the correlation length of the 2D component ($R_g < L_{c,2D}$), and therefore they are efficiently trapped in cross-field 2D structures (see Figures 1(b), and 1(c)), thus inhibiting parallel diffusion. This implies that QLT applied to composite turbulence breaks down at low rigidities because the assumption of a nearly undisturbed particle gyro-orbit needs to be replaced by a disturbed orbit approach such is the case in a fully nonlinear theories (e.g., Shalchi et al. 2004, le Roux & Webb 2007, Qin 2007, Yan & Lazarian 2008).

The perpendicular diffusion ($\kappa_\perp$) coefficient is shown in Figure 2(b) for the same turbulence models and rigidities as in Figure 2(a). Here $\kappa_\perp$ is defined as $\kappa_\perp = <(\Delta r_\perp)^2>/(2t)$, where the ensemble averaging is the same as for $\kappa_\parallel$. As expected, the pure slab model yields the well-known subdiffusive behavior, $<(\Delta r_\perp)^2> \sim t^{0.5}$ asymptotically at late times when particles are diffusing along the random-walking field lines but staying on the same field line (e.g., Qin et al. 2002a). This inhibition of cross-field transport is a direct consequence of the limitations particles experience when propagating in non-3D turbulence where only slab ($k_\parallel$) wave modes are present (Jones et al. 1998). Cross-field diffusion is recovered for composite turbulence (quasi-3D turbulence) at late times (e.g., Qin et al. 2002b, Giacalone & Jokipii 1999) and it is approximately proportional to the 2D turbulence component energy. This stems from particles that are transported across the mean magnetic field more efficiently as they propagate along field lines that separate more rapidly as field-line wandering becomes more pronounced with enhanced 2D turbulence energy. Particle trapping by 2D loop-like structures (1st panel of Figure 1(b)) does not appear to hinder perpendicular transport more with increasing energy of the 2D turbulence component when the structures are stronger. This suggests that there are not enough such structures to have strong impact (see first panel of Figure 1(a)). Particle trapping will even be less significant at high rigidities when $R_g > L_{c,2D}$ because trapping only occurs when $R_g < L_{c,2D}$ (e.g., Verkhoglyadova & le Roux 2005).

The rigidity dependency of the parallel and perpendicular mean free paths is shown in Figure 3 based on the definitions $\lambda_{\parallel,\perp} = 3\kappa_{\parallel,\perp}/v$. Figure 3(a) shows the simulated parallel mean free path as a function of particle rigidity for slab and composite turbulence models. The results are compared with the QLT prediction of Jokipii (1966) for slab turbulence. The QLT parallel mean free path for slab turbulence can be expressed as (see Equation 3.39 in Shalchi 2009):

$$\lambda_{\parallel,QLT} = \frac{3L_{sb}}{8\pi C} \frac{B_0^2}{\delta b_{sb}^2} R^{2-2\nu} \left[ \frac{1}{1-\nu} {}_2F_1\left(1-\nu,-\nu,2-\nu,-R^2\right) - \frac{1}{2-\nu} {}_2F_1\left(2-\nu,-\nu,3-\nu,-R^2\right) \right] \quad (11)$$

In this equation $L_{sb}$ is the slab bend-over scale, which is linked to the correlation length by $L_{c,sb} = 2\pi C L_{sb}$, $\nu$ is half the Kolmogorov power-law slope index $\gamma = 5/3$, $R$ is the normalized rigidity (gyroradius normalized to the turbulence correlation scale), $C$ is the power spectrum normalizing constant $C(\gamma,s) = \Gamma((\gamma+s)/2)/[2\Gamma((\gamma-1)/2)\Gamma((s+1)/2)]$, and $s$ is the energy-containing range power-law index. Even though we assumed strong turbulence, the results in Figure 3(a) confirm that QLT, valid strictly speaking only for weak turbulence, is a good approximation for $\lambda_\parallel$ in the case of slab geometry. QLT was shown (e.g., Bieber et al. 1994, le Roux et al. 2004, Shalchi & Shlickeiser 2004), however, to be non-valid for any non-slab static turbulence model. This includes composite and isotropic static turbulence models, both of which yield an infinite value for $\lambda_\parallel$. This divergence is a consequence of the vanishing Fokker-Planck pitch angle coefficient ($D_{\mu\mu}(\mu = 0) = 0$). The reported comparisons with QLT in Figure 3 for the case of composite turbulence was done by showing the QLT value for $\lambda_\parallel$ for the slab portion of the composite turbulence only, thus avoiding the problem of infinite values for $\lambda_\parallel$ produced by QLT for static 2D turbulence.

The results indicate that the rigidity dependence of $\lambda_\parallel$ for composite turbulence is considerably stronger than predicted by QLT for slab turbulence, and that this dependence becomes stronger the more 2D turbulence dominates composite turbulence. Thus, $\lambda_\parallel$ is strongly reduced by 2D turbulence at low rigidities where $R_g < L_{c,2D}$, but has basically no effect at high rigidities since the effect of slab turbulence dominates when $R_g > L_{c,2D}$ as discussed above. QLT for slab turbulence combined with the prediction of negligible particle scattering by QLT for 2D turbulence seem to qualitatively explain the results for $\lambda_\parallel$ at high rigidities, but at low rigidities a nonlinear (disturbed gyro-orbit) approach where parallel and perpendicular diffusion are coupled shows more promise. Such approaches can in fact qualitatively explain the increased rigidity dependence of $\lambda_\parallel$ in composite turbulence when perpendicular diffusion dominated by 2D turbulence feeds back into parallel diffusion as shown in the formulation of weakly non-linear theory (WNLT) (Shalchi et al. 2004, le Roux & Webb 2007) and extended non-linear guiding centre theory (ENLGC) (Qin 2007).

In Figure 3(b) we present the simulated perpendicular mean free path $\lambda_\perp$ for slab and composite turbulence models and compare it with well-known theoretical models for perpendicular diffusion. The dotted lines indicate the results for the field-line random walk (FLRW) theory for perpendicular diffusion of fast particles in composite turbulence. To calculate the FLRW perpendicular diffusion coefficient we used the expression of Matthaeus et al. (1995) for the field-line random walk of composite slab and 2D turbulence which yields a perpendicular mean free path given by

$$\lambda_{\perp,FLRW} = \frac{3\pi}{2B_0^2}\left[\delta b_{sb}^2 L_{c,sb} + \sqrt{\left(\delta b_{sb}^2 L_{sb}\right)^2 + 4\left(\frac{\delta b_{2d}^2 L_{c,2D}}{2\sqrt{6}}\right)^2}\right] \quad (12)$$

The dotted lines are independent of particle rigidity because, as the expression indicates, $\lambda_\perp$ depends only on the characteristics of the field-line-random-walk process (ballistic particle motion along the turbulent magnetic field). The FLRW result for slab turbulence can also be derived using a QLT kinetic approach for energetic particle transport (e.g., Jokipii 1966, Forman et al. 1974). The FLRW model provides the most efficient means for particle transport across the mean magnetic field because particles follow random-walking field lines freely without scattering ($\lambda_\parallel \gg L_c$) as was noted before (e.g., Qin et al., 2002a, b, Giacalone & Jokipii, 1999). Note that when 2D turbulence dominates FLRW (Matthaeus et al., 1995), perpendicular FLRW diffusion is reduced by particle trapping in 2D structures.

However, the simulations, which are valid for strong turbulence, predict much less perpendicular diffusion than indicated by the FLRW perpendicular diffusion model (particles at late times are diffusing along wandering field lines instead of going along them ballistically). We find that the simulations are well reproduced by non-linear guiding center theory (NLGC) theory (Matthaeus et al. 2003, Shalchi et al. 2008, Zank et al. 2004, le Roux & Webb 2007) when 2D turbulence is dominant assuming that the free parameter in the theory $a = 1$. The NLGC theory formulation for $\lambda_\perp$ (Matthaeus et al. 2003) is:

$$\lambda_{\perp,NLGC} = \frac{2a^2 C}{B_0^2} \lambda_\| \left[ \delta b_{sb}^2 L_{sb} \int_0^\infty \frac{(1+k_\|^2 L_{c,sb}^2)^{-\nu}}{1+(k_\|^2 \lambda_\|^2/3)} dk_\| + \delta b_{2d}^2 L_{2d} \int_0^\infty \frac{(1+k_\perp^2 L_{c,2D}^2)^{-\nu}}{1+(k_\perp^2 \lambda_\| \lambda_\perp/3)} dk_\perp \right] \quad (13)$$

where $\lambda_\|$ is specified using our simulations. A key element in the success of this model is that the undisturbed particle orbit along a uniform magnetic field used in QLT is replaced by guiding center motion that is diffusive because guiding centers follow the actual random-walking turbulent magnetic field and their motion are also affected by pitch-angle scattering from gyro-resonant interaction with micro-scale slab turbulence ($\lambda_\| \gg L_c$). In contrast, QLT for dominant static 2D turbulence predicts $\lambda_\perp$ to be infinite to indicate superdiffusion (Shalchi & Schlickeiser, 2004) and thus fails to reproduce our simulations and published results involving other simulations (e.g., Qin et al. 2002a,b; Giacalone & Jokipii 1999). In the simulations for composite turbulence the rigidity dependence in $\lambda_\perp$ is positive but weak which is well reproduced by NLGC theory based on numerical calculations using Equation (13) as shown in Figure 3(b). Analytical estimates of the rigidity dependence suggest that $\lambda_\perp \propto R^{1/9}$ (e.g., Shalchi & Schlickeiser 2004, Zank et al. 2004). On the downside, NLGC theory predicts diffusion instead of subdiffusion in the limit of slab turbulence, thus failing to reproduce the subdiffusive simulation results. However, a modified version of NLGC theory, the so-called extended NLGC theory (ENLGC) by Shalchi (2006) addressed this limitation. Note that the simulated rigidity dependence of $\lambda_\perp$ for slab turbulence is negative. This is consistent with the rigidity dependence predicted theoretically for compound diffusion (subdiffusion) (e.g., Chuvilgin and Ptuskin 1993) when particles are diffusive along and tied to random-walking field lines, whereby $\lambda_\perp \propto R^{-1/3}$ for slab turbulence at non-relativistic energies.

NLGC theory includes the effect of parallel diffusion on perpendicular diffusion but ignores the effect of perpendicular diffusion on parallel transport since the focus was on perpendicular diffusion. In more general theories a two-way feedback between parallel and perpendicular transport is included in the formulation of weakly non-linear theory (WNLT) (Shalchi et al. 2004, le Roux & Webb 2007) and in the extended NLGC theory of Qin (2007). These approaches improved the description of the rigidity dependence of $\lambda_\parallel$ in the case when 2D turbulence is dominant and the perpendicular diffusion due to 2D turbulence has a strong feedback effect on parallel transport.

### 3.2. Intermittent turbulence

To generate an intermittent $q$-Gaussian magnetic field we use the procedure described in Section 2, Equations (7)-(9). Figures 4(a), 4(b), and 4(c) show the parallel and perpendicular diffusion coefficient as a function of time for particle rigidity $R = 0.075$ ($R_g/L_c = 0.075$). Figure 4(a) compares uniform and intermittent slab turbulence. The latter is somewhat similar, in absence of dynamical effects and magnetic field compressibility, to multi-scale Afvénic turbulence (Beresnyak et al. 2011). The intermittency parameters ($q$, $L_P$) are (1.55, $0.1L_{c,sb}$) and (1.27, $3.3L_{c,sb}$) respectively.

Overall, one sees that parallel transport in intermittent slab turbulence can be considerably less efficient compared to parallel transport in uniform slab turbulence. Since there is statistically a high probability of particles encountering intermittent regions of weak slab turbulence and a low probability, but much higher than in the uniform case, of encountering regions of strong turbulence, it indicates that particles must spend considerable time in regions of strong turbulence to explain the large reduction in parallel transport when turbulence is intermittent.

Unlike the uniform slab turbulence case where the parallel diffusion coefficient's initial superdiffusive phase is followed by a diffusive phase at late times, the parallel diffusion coefficient for intermittent slab turbulence (Figure 4(a)) shows an extended subdiffusive phase after an even shorter-lived initial superdiffusive phase before reaching diffusion at late times. The subdiffusive phase can come about when particles enter patches of strong turbulence in the $z$-direction for the first time after propagating in regions of weak turbulence. This sudden change in the turbulence level causes an increase in the pitch-angle scattering rate resulting in a temporary slowdown in parallel diffusion with time (subdiffusion). We also find that the parallel diffusion coefficient is larger for smaller-scale intermittent turbulence regions. One could have argued that particles would encounter more patches of strong turbulence during the same time if the patches are smaller, thus statistically better sampling the tail of the magnetic field PDF, experiencing enhanced scattering more frequently, and consequently have a reduced parallel diffusion coefficient compared to the large-scale intermittency case as discussed in Alouani-Bibi & le Roux (2012). The fact that this did not happen indicates that what is more important is the average residence time of particles in strong turbulence patches. For the small intermittency patch size case (small $L_p$-value), most of the particles considered here have a gyroradius of the same order or larger than the intermittency patch size. Therefore particles will spend less time within a patch of strong turbulence. For the larger $L_p$-value, on the other hand, even though the tail of the magnetic field turbulence PDF is statistically not well sampled due to a small number of patches in the simulation box, $L_p$ is much larger than the gyroradius for all particles considered. Particles will therefore spend a long time inside a patch of strong turbulence. Thus, even though they will encounter only a few strong turbulence patches, a stronger longer-lasting reduction in the parallel diffusion coefficient (longer-lasting subdiffusive phase) of particles is then the outcome.

Because of the multi-scale property of observed intermittent turbulence in the solar wind, one may expect such a subdiffusive phase to be a general feature of parallel transport in solar wind turbulence conditions. One might also expect that this feature will be stronger and more long-lasting for larger values of $L_P$ that exist in the solar wind, but such large values were not addressed in our simulations. This effect, however, will only extend to turbulence scales small enough where the PDF is non-Gaussian ($q > 1$), because for Gaussian turbulence the probability of particles encountering a patch of strong turbulence decreases exponentially as $e^{-x^2}$.

We also investigated the duration of the subdiffusive phase for parallel transport at lower rigidities (the results are not shown). For the large-scale intermittent turbulence patches (large $L_p$-value), low rigidity particles ($R = 0.025$) showed a more long-lasting subdiffusive phase (longer residency time in strong turbulence patches because lower-energy particles have a shorter parallel scattering mean-free path). This is supported by the estimate of the slope $s$ for the mean-square particle displacement $<(\Delta z)^2> \sim t^s$ at the end of the calculation which was s = 0.97 (traces of subdiffusion still remains at late times). For high rigidity particles with $R = 0.25$, the subdiffusive phase was of short duration which is consistent with an increased value of $s = 0.99$ indicating a more complete diffusive state at the end of the calculation. These results are consistent with the argument that the residency time of particles in patches of strong turbulence explains the results best.

The most unusual feature in Figure 4(a), is that for large-scale intermittency ($L_P = 3.3L_{c,sb}$) we see that $\kappa_\perp / \kappa_\parallel > 1$ until late times for a duration of about $\sim 4 \times 10^4 \Omega^{-1}$. A similar effect is seen when comparing cross-field diffusion for $L_P = 3.3L_{c,sb}$ to parallel diffusion for $L_P = 0.1L_{c,sb}$. in this case $\kappa_{\perp, L_P=3.3Lc} / \kappa_{\parallel, L_P=0.1Lc} > 1$ again, but this occurs for a period $\sim 3 \times 10^3 \Omega^{-1}$ that is ten times shorter. This unusual high level of perpendicular transport relative to parallel transport can be understood if particles after the initial superdiffusive phase get trapped in a few patches of strong turbulence. Particle transport along the turbulent magnetic field is strongly reduced in such a patch as discussed above, which suggests that subdiffusion should also be reduced because for this compound perpendicular diffusion process (e.g., Chuvilgin and Ptuskin 1993, Webb et al. 2006, 2009) where fast particles diffuse along a random-walking field line $\kappa_\perp \propto (\delta b_{sb} / B_0)^2 \sqrt{\kappa_\parallel / t}$ (compound diffusion theory strictly speaking does not apply to such strong levels of turbulence referred to here but is useful as a guide to interpret the results). However, the strong increase in the slab turbulence level inside the intermittent patch will boost the efficiency of compound diffusion according to the theory. Our results indicate that the increased level of field-line random walk dominates the reduced level of parallel diffusion in determining the enhanced compound diffusion level in the simulations for large-scale patches. This is consistent with the theoretical prediction that compound diffusion is more weakly dependent on parallel diffusion than on the field-line-random walk turbulence level. Note that for small-scale intermittent turbulence patches the compound diffusion level is found to be comparable with the uniform turbulence case.

The simulation results for intermittent composite turbulence are shown in Figure 4(b) where the slab and 2D components have the same energy, and in Figure 4(c) where the 2D component is dominant. The most striking aspect of adding intermittency to composite turbulence is that the subdiffusive phase for parallel transport has a considerably steeper negative slope compared to intermittent slab turbulence in the case of the large-scale intermittent turbulence regions. This feature is at its strongest in Figure 4b where the slab and 2D components have equal energy. This can be understood as a consequence of both slab and 2D turbulence slowing down parallel transport when particles enter a strong turbulence patch for the first time. There is a slowdown in parallel transport (subdiffusion) because of stronger pitch-angle scattering through gyro-resonant interaction with enhanced slab turbulence in the high turbulence region. This slowdown, however, is diminished to some extent by the overall decrease in the slab turbulence when considering composite turbulence. Combined with this slowdown is the additional more potent slowdown that follows because particles are temporarily trapped by strong quasi-coherent 2D structures in the 2D turbulence component when entering the highly turbulent region. However, given the 3D nature of composite turbulence, particles are free to cross field lines in contrast to the slab turbulence case (e.g., Jones et al. 1998). This results in the restoration of parallel diffusive particle behavior at late times as particles escape from the large turbulence patch and the probability of encountering another strong large turbulence patch is low (only a few large turbulence patches can be encountered in the simulation box). Analysis of the parallel diffusion coefficient slope $s$ for composite turbulence showed a slightly stronger rigidity dependence than for intermittent slab turbulence. The values for *s* at the end of the simulations are as low as ~0.85 for $R = 0.025$ if $\delta B^2_{2D}/\delta B^2_{sb} = 1$, and ~0.94 if $\delta B^2_{2D}/\delta B^2_{sb} = 4$ (subdiffusion), while $s \approx 1$ when $R = 0.25$ (diffusion).

For perpendicular transport in composite turbulence the effect of intermittency is much smaller and more subtle. Perpendicular transport is reduced for most cases but enhancement is also possible (see Figure 4(b)). Concerning composite turbulence, perpendicular transport during intermediate times is considerably less subdiffusive and the subdiffusive phase is of shorter duration when 2D turbulence dominates slab turbulence and intermittency is large-scale (Figure 4(c)). Thus the late-time recovery of classical perpendicular diffusion is reached much earlier because of the prominent role of the 2D turbulence component in transporting particles across the mean magnetic field through field-line random walk when particles encounter one or a few large-scale regions of enhanced turbulence.

Figure 5(a) shows the rigidity dependence of the parallel mean free path for slab and composite turbulence geometries including both uniform and intermittent turbulence. As was the case in Figure 3, the mean free paths are calculated using the spatial diffusion coefficients derived from our simulation results according to $\lambda_\parallel = 3\kappa_\parallel/v$. In the QLT limit, $\lambda_\parallel$ can be derived by integrating over-all pitch angles the function $(1-\mu^2)^2/D_{\mu\mu}(\mu)$, where $D_{\mu\mu}$ is the pitch-angle Fokker-Planck coefficient (e.g., Schlickeiser 1989, Shalchi 2009). The reason for not using the later approach is the difficulty of building up enough statistics in the pitch-angle bins to numerically integrate with some accuracy $D_{\mu\mu}(\mu)$ when the turbulence is strong ($\delta B/B_0 \sim 1$) as pointed out in Qin & Shalchi (2009). Since the comparison between parallel mean free paths for uniform slab and composite turbulence has already been presented in Figure 3(a) above, the focus here will be on comparing results for the intermittent slab and 2D turbulence cases.

The parallel mean free path for intermittent turbulence is smaller than in the uniform turbulence case over a wide range of rigidities (Figure 5(a)). For slab turbulence this is due to particles encountering many regions of strong turbulence and thus experiencing frequent episodes of strong pitch-angle scattering along the background magnetic field for small-scale intermittency. In the case of large-scale intermittency, the main inhibitor of strong parallel transport is the long-lasting trapping of particles in a few strong turbulence patches where strong pitch-angle scattering occurs for much longer periods. More energetic particles than what is considered in this paper, i.e., those with $R_g > 3.3 L_{c,sb}$, will have too large gyroradii for efficient trapping in patches of strong turbulence to occur. In this case particles sample strong turbulence patches as scattering centers which happens more frequently given the absence of particle trapping. Consequently, parallel transport is reduced compared to the uniform slab turbulence. For the intermittent composite turbulence model, the reduced $\lambda_\parallel$ is mainly the result of a much stronger 2D component, causing strong particle confinement within the perpendicular plane.

The rigidity dependence of the ratio $\lambda_\perp/\lambda_\parallel$ in Figure 5(b) for intermittent composite turbulence has similar features as in the uniform turbulence case as was previously reported (Shalchi et al. 2004, Tautz et al. 2006, Tautz & Shalchi 2010, Qin & Shalchi 2012). The case of uniform turbulence and turbulence with intermittent small-scale patches are similar in that they both show a diminishing ratio for $\lambda_\perp/\lambda_\parallel$ with increasing particle rigidity. This behavior is theoretically expected for both the slab and composite turbulence model assuming uniform turbulence. For instance, according to classical NLGC theory for composite turbulence dominated by 2D turbulence $\lambda_\perp \propto R^{1/9}$ while $\lambda_\parallel \propto R^{1/3}$ following the QLT prescription assuming that slab turbulence is more important for parallel transport than 2D turbulence (e.g., Zank et al. 2004). This would imply that $\lambda_\perp/\lambda_\parallel \propto R^{-2/9}$ yielding a weak decrease with increasing rigidity, which is consistent with the simulations. In the limit of slab turbulence, the ratio $\lambda_\perp/\lambda_\parallel$ decreases considerably more strongly with increasing rigidity for both the uniform and intermittent turbulence cases. This is due to the subdiffusive (compound diffusion) behavior of perpendicular transport according to which $\kappa_\perp \sim (\kappa_\parallel)^{1/2}$ so that $\lambda_\perp/\lambda_\parallel \propto 1/(\kappa_\parallel)^{1/2}$. According to QLT for slab turbulence $\kappa_\parallel \propto R^{4/3}$ so that $\lambda_\perp/\lambda_\parallel \propto R^{-2/3}$ which theoretically explains the stronger negative rigidity dependence of this ratio in the case of slab turbulence.

For large-scale intermittency the ratio $\lambda_\perp/\lambda_\parallel$ is of the order one for the composite turbulence at all rigidities shown. This is mainly caused by the strong reduction in the efficiency of parallel transport as particles enter one or few large regions with strongly enhanced composite turbulence. A combination of particle trapping by 2D structures, field-line random walk, and enhanced particle scattering by the slab turbulence in the strong composite turbulence region severely curtails parallel transport.

In the slab turbulence case, even though the ratio $\lambda_\perp/\lambda_\parallel$ decreases with increasing rigidity as explained above, the ratio remains of the order of one because of the strong impedance of parallel transport in large patches of enhanced slab turbulence where particles are strongly scattered and field-line random walk reduces parallel transport, but also because of the significant increase in perpendicular subdiffusion in these patches due to enhanced field-lined random walk. The ratio will become increasingly smaller so that $\lambda_\perp/\lambda_\parallel < 1$ if the simulation is continued for a longer time as perpendicular subdiffusion continues to suppress perpendicular diffusion further ($\kappa_\perp \propto 1/t^{1/2}$). However, one needs to keep in mind that at later times the validity of static turbulence assumption may be jeopardized.

To assess the effect of intermittent turbulence on higher moments of particle displacement, we estimated the kurtosis for particles of a given rigidity. Only moments along the background magnetic field are considered in deriving the kurtosis because turbulence is specified to be intermittent only in this direction. For each moment the ensemble averaging is performed in the same way as was done for the diffusion coefficients (second moments). Figures 6(a), 6(b), and 6(c) present the time-dependent kurtosis for the slab and composite turbulence models for $R = 0.075$. The kurtosis ($K$) mimics the behavior of the diffusion coefficient in that it is initially far from the Gaussian limit ($K = 3$) and then converges towards the Gaussian value at late times. In the case of uniform turbulence the Gaussian value is reached after $\sim 2.5 \times 10^4$ gyro-periods. However, in the case of intermittent composite turbulence, the kurtosis converged to values between 3-3.5 at late times with the largest deviation from the Gaussian limit coming from the case of large-scale intermittency. Smaller deviations from the Gaussian limit also appear for intermittent slab turbulence. Such deviations are most prominent for particles with low rigidities because these particles sample small-scale magnetic field fluctuations whose PDF profiles most strongly display non-Gaussian characteristics (larger $q$-value). Thus, kurtosis tells us that whereas parallel particle transport is diffusive in uniform static turbulence at late times, the diffusive limit is never reached in the case of intermittent static turbulence.

## 4. Conclusions

In our energetic particle trajectory simulations we included intermittency in the magnitude of the turbulent magnetic field along the mean magnetic field in our idealized model of solar wind turbulence which consists of a combination of a slab and a 2D component. Using the non-Gaussian statistical characteristics (the entropic index $q$ and intermittency patch size, derived from observations), we generated an intermittent quasi-3D model of turbulence. Energetic particle diffusion coefficients both parallel and perpendicular to the mean field were calculated from the test-particle simulations in this composite turbulence model for two intermittency scales. These diffusion coefficients are found to be significantly affected by the intermittency of the turbulence. The effect is most significantly present for parallel transport. Parallel diffusion is strongly reduced by the presence of intermittency, especially in the case of composite turbulence with large-scale intermittency where particles interact with one or at most a few large-scale regions with strongly enhanced turbulence. In this case an extended period of parallel subdiffusion after the initial superdiffusive phase develops during intermediate times before diffusion is recovered at late times. In the case of uniform turbulence the initial superdiffusive phase is followed directly by diffusion.

For perpendicular transport the effect of intermittency is typically much smaller and more subtle (the exception is in the limit of slab turbulence where perpendicular subdiffusion is considerably enhanced when intermittency is large-scale). Perpendicular transport in composite intermittent turbulence is reduced in most cases but enhancement is also possible. Concerning composite turbulence, perpendicular transport during intermediate times is considerably less subdiffusive and the subdiffusive phase is of shorter duration when 2D turbulence dominates slab turbulence and intermittency is large-scale. Thus the late-time recovery of classical perpendicular diffusion is reached much earlier because of the prominent role of the 2D turbulence component when particles encounter and are trapped in one or only a few large-scale regions of enhanced turbulence.

It is interesting to note that the ratio of perpendicular diffusion over parallel diffusion is enhanced in the presence of intermittent turbulence at all particle rigidities studied. This enhancement is such that the ratio reaches values of the order of one when assuming large-scale intermittent turbulence (perpendicular transport is more efficient than parallel transport). Similar large ratios were determined from observations at corotating interaction regions by Dwyer et al. (1997) which included a maximum ratio of ~1.5. However, in their case they explained the large ratio as due to enhanced perpendicular transport whereas here we found the reason to be strongly reduced parallel transport. If the parallel diffusion coefficient in actual intermittent solar wind turbulence is reduced to the extent suggested by our simulations, this result will have significant implications for our understanding of solar energetic transport in the inner heliosphere as well as for cosmic-ray modulation in the large-scale heliosphere.

An investigation of the higher order moments of parallel transport in terms of the kurtosis revealed that when turbulence is intermittent, particles at lower rigidities never quite reach the diffusive limit (kurtosis $K = 3$) at late times in contrast to the uniform turbulence case. Although the kurtosis values at late times are not far from three, this indicates that some of the non-Gaussian statistical properties of the intermittent turbulent magnetic field are transmitted to low rigidity particles. These results indicate that higher order moments of particle displacement must be taken into account in order to achieve an accurate description of energetic charged particle transport in intermittent turbulence. Thus, parallel transport theories for energetic particles, such as QLT, that solely focus on the contribution of the second moment of particle displacement might need revision to include higher order moment contributions. This implies that particle transport theory needs to be developed to include a generalized statistical description of the turbulent interplanetary magnetic field that is non-Gaussian and is expanded to include higher order turbulence correlation functions.

## Acknowledgments

This work was supported by NSF-DOE grant ATM-0904007 and by NASA grant NNX10AC15G. F. A.-B. was partially supported by the NASA grants NNX10AE46G, NNX11AB48G, NNX12AH44G. The publication charges for this manuscript were supported by the CSPAR and the Dept. of Space Science, University of Alabama in Huntsville.

**Figure Captions**

Figure 1. (a) Magnetic field lines along the $z$-axis, the axis of the background magnetic field for composite turbulence, with $\delta B^2_{2D}/\delta B^2_{sb} = 4$. The left most panel is for uniform turbulence, middle panel is for intermittent turbulence with entropic index $q = 1.55$ and small-scale intermittency patch size $L_P = 0.1 L_{c,sb}$, while the right most panel is for intermittent turbulence with $q = 1.27$ and large-scale patch size $L_P = 3.3 L_{c,sb}$. (b) Magnetic field lines projected in the transverse $xy$-plane perpendicular to the background magnetic field. (c) The magnetic field vector along the $xy$-plane for composite uniform turbulence. (d) The same as (a) but valid for slab turbulence.

Figure 2. (a) The parallel diffusion coefficient $\kappa_\parallel$, derived from particle trajectory calculations, in units of $10^{23}$ cm$^2$/s as a function of time in units of the particle gyroperiod associated with the mean magnetic field. (b) The corresponding perpendicular diffusion coefficient $\kappa_\perp$ as function of time for uniform turbulence. Three models of turbulence are shown: Squares for slab turbulence, circles for composite turbulence (slab + 2D), with $\delta B^2_{2D}/\delta B^2_{sb} = 1$, and triangles for composite turbulence (slab + 2D) with $\delta B^2_{2D}/\delta B^2_{sb} = 4$. For all cases in all figures $\delta B^2_{tot}/B_0^2 = 1$. Filled symbols represent particles with a normalized rigidity of $R = 0.025$ (ratio of the particle gyroradius over the slab turbulence correlation length), while open symbols are for particles with $R = 0.25$.

Figure 3. (a) The parallel mean free path $\lambda_\parallel$, derived from particle trajectory calculations, as function of normalized rigidity $R$ for uniform turbulence. Three models of turbulence are shown: A solid line with squares for slab turbulence, a solid line with circles for composite turbulence (slab + 2D) with $\delta B^2_{2D}/\delta B^2_{sb} = 1$, and a solid line with triangles for composite turbulence (slab + 2D) with $\delta B^2_{2D}/\delta B^2_{sb} = 4$. The dashed lines represent the results of quasi-linear theory (QLT) for slab turbulence. The symbols on dashed lines indicate the fraction of the total turbulence energy assumed to be in the slab component for input to the QLT calculation: Squares - 100%, circles – 50%, and triangles – 20%. Both the quantities on the *x*- and the *y*-axis are normalized to the slab turbulence correlation length of 0.03 AU. (b) The corresponding perpendicular mean free path $\lambda_\perp$ as function of $R$ for uniform turbulence. Three models of turbulence are shown: Solid line with squares – slab turbulence, solid line with circles – composite turbulence (slab + 2D) with $\delta B^2_{2D}/\delta B^2_{sb} = 1$, and solid line with triangles - composite turbulence (slab + 2D) with $\delta B^2_{2D}/\delta B^2_{sb} = 4$. The dashed lines represent the results of non-linear guiding center theory (NLGC) for $\lambda_\perp$

with $\lambda_\parallel$ taken from our particle trajectory simulations. Dotted lines represent the results for $\lambda_\perp$ based on field line random walk theory (FLRW). The symbols on dashed or dotted lines indicate the assumed fraction of turbulence energy in the slab component used to determine $\lambda_\perp$ theoretically: Squares - 100%, circles - 50%, and triangles - 20%. Both the *x*- and the *y*-axis are normalized to the slab turbulence correlation length assumed to be 0.03 AU.

Figure 4. The parallel diffusion coefficient $\kappa_\parallel$ (solid lines with filled symbols) and perpendicular diffusion coefficient $\kappa_\perp$ (dashed lines with open symbols) as function of time as determined from particle trajectory simulations. Results are shown for the normalized particle rigidity $R = 0.075$. Squares indicate uniform turbulence, circles denote intermittent turbulence along the mean magnetic field with entropic index $q = 1.55$ and small-scale intermittency patch size $L_P = 0.1 L_{c,sb}$, and triangles imply intermittent turbulence with entropic index $q = 1.27$ and large-scale intermittency patch size $L_P = 3.3 L_{c,sb}$. (a). Results for slab turbulence. (b). Results for composite turbulence (slab + 2D) with $\delta B^2_{2D}/\delta B^2_{sb} = 1$. (c). Results for composite turbulence (slab + 2D) with $\delta B^2_{2D}/\delta B^2_{sb} = 4$.

Figure 5. (a) The normalized parallel mean free path $\lambda_\parallel$ as function of normalized particle rigidity $R$. (b) The corresponding ratio of the perpendicular to the parallel mean free path ($\lambda_\perp/\lambda_\parallel$) as function of $R$. Solid lines are for slab turbulence, dashed lines represent composite turbulence (slab + 2D) with $\delta B^2_{2D}/\delta B^2_{sb} = 1$, and dotted lines indicate composite turbulence (slab + 2D) with $\delta B^2_{2D}/\delta B^2_{sb} = 4$. Squares are valid for uniform turbulence, circles for intermittent turbulence with entropic index $q = 1.55$ and small-scale intermittency patch size $L_P = 0.1 L_{c,sb}$, and triangles for intermittent turbulence with $q = 1.27$ and large-scale patch size $L_P = 3.3 L_{c,sb}$.

Figure 6. The kurtosis of parallel particle transport as function of time. Results are shown for $R = 0.075$. Squares are valid for uniform turbulence, circles denote intermittent turbulence with entropic index $q = 1.55$ and small-scale intermittency patch size $L_P = 0.1 L_{c,sb}$, and triangles represent intermittent turbulence with entropic index $q = 1.27$ and large-scale intermittency patch size $L_P = 3.3 L_{c,sb}$. The dashed line represent the kurtosis value $K = 3$ for particles obeying Gaussian statistics along the mean magnetic field. (a) Results for slab turbulence. (b) Results for composite turbulence (slab + 2D) turbulence with $\delta B^2_{2D}/\delta B^2_{sb} = 1$. (c) Results for composite turbulence (slab + 2D) with $\delta B^2_{2D}/\delta B^2_{sb} = 4$.

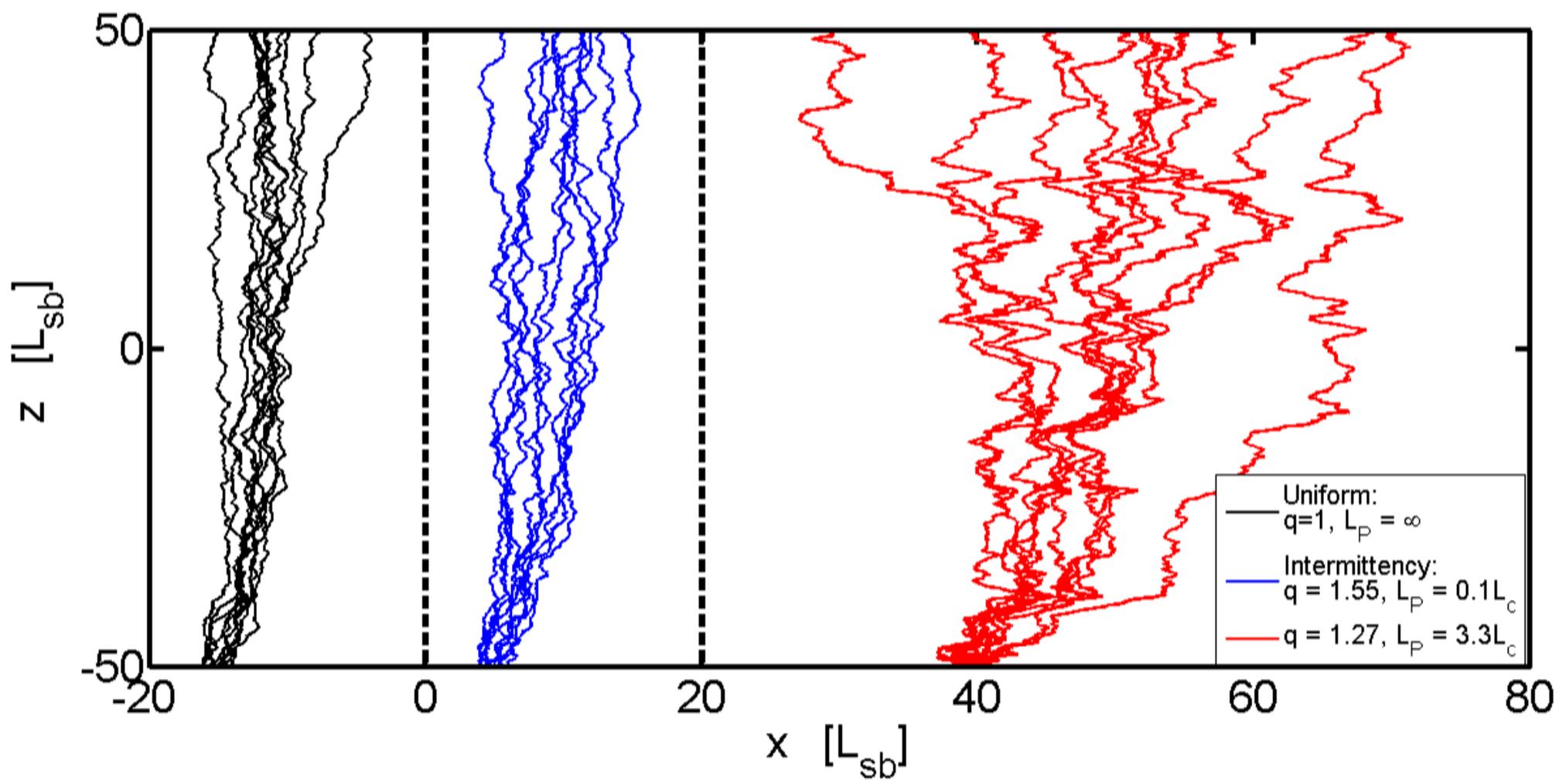

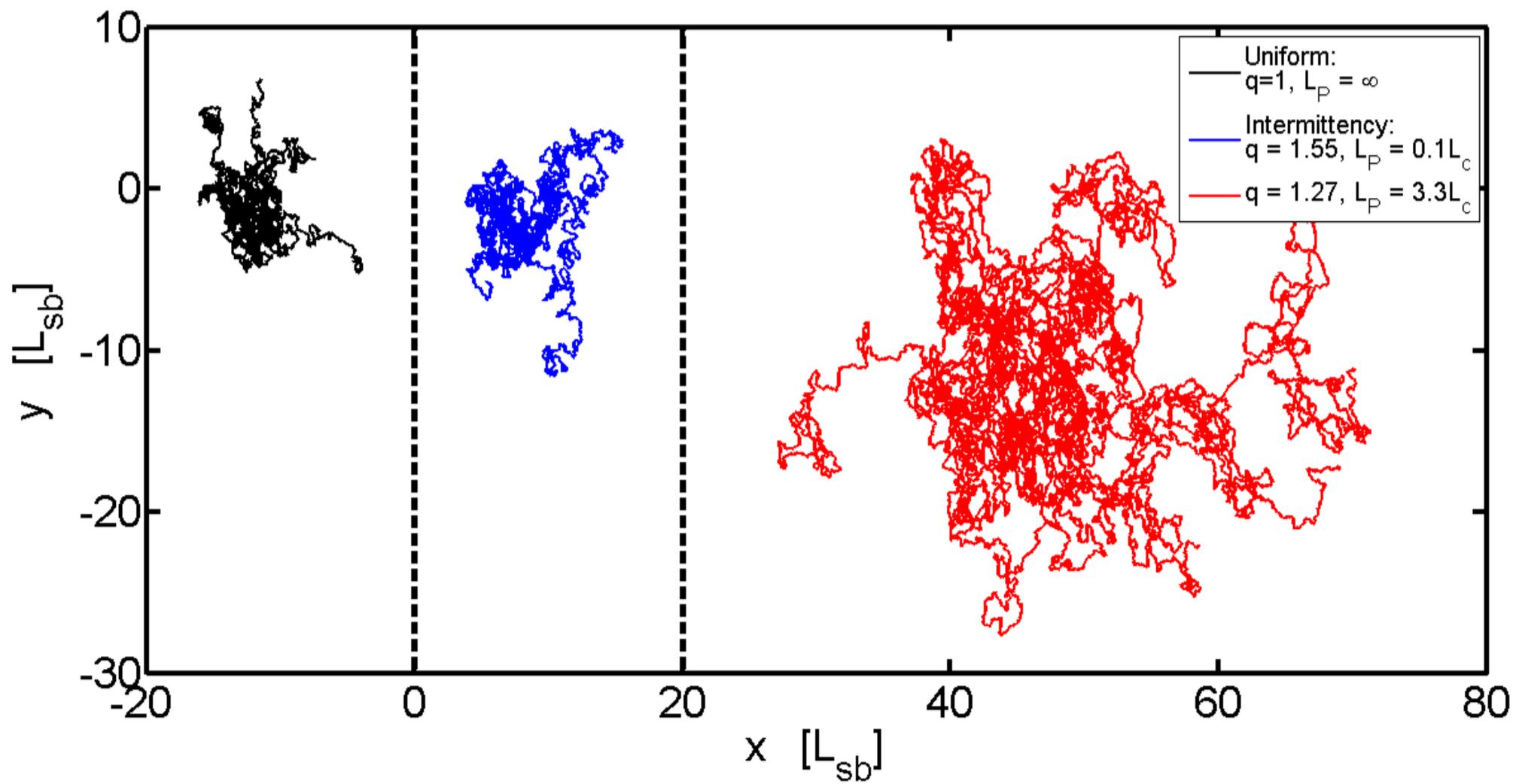

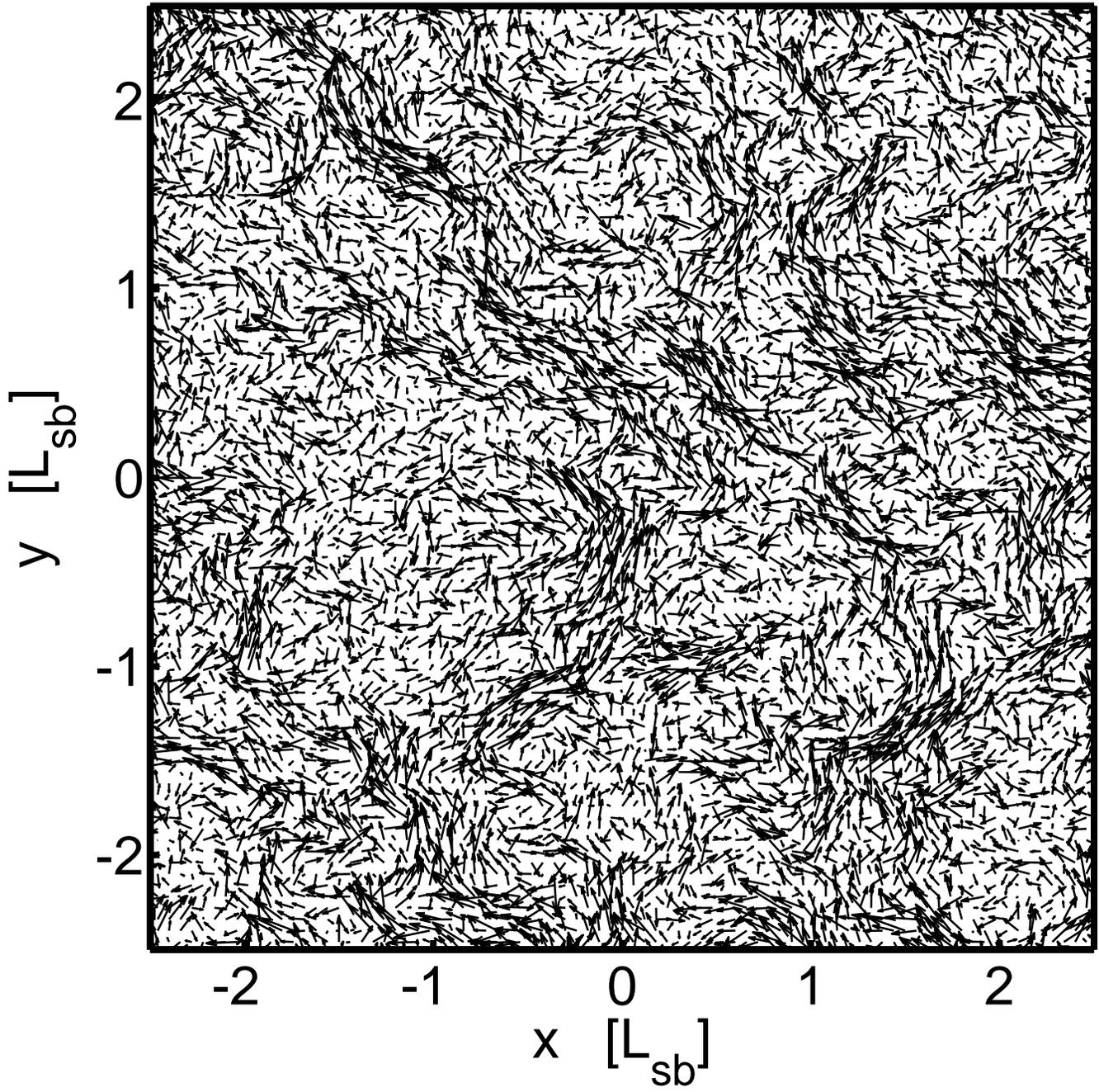

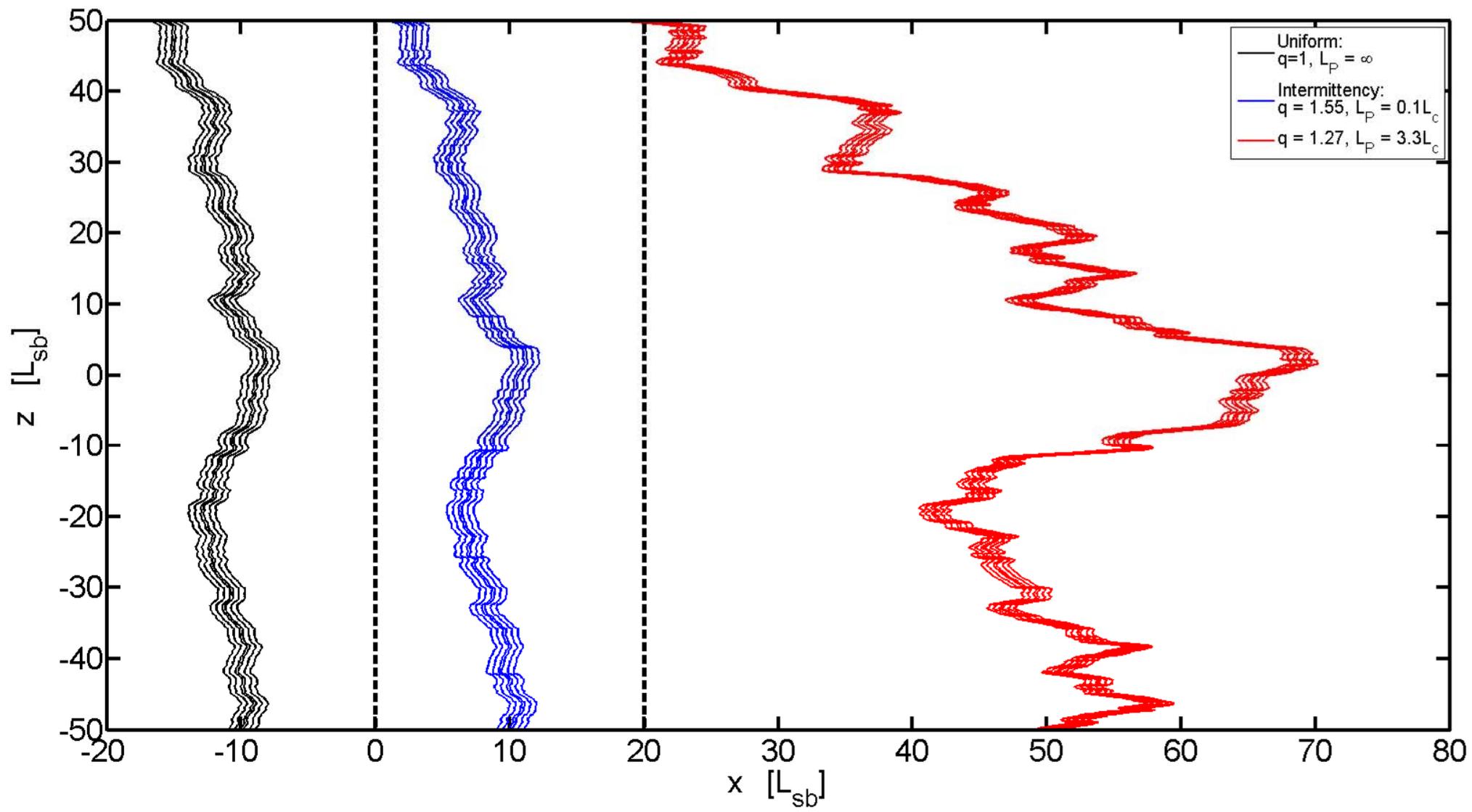

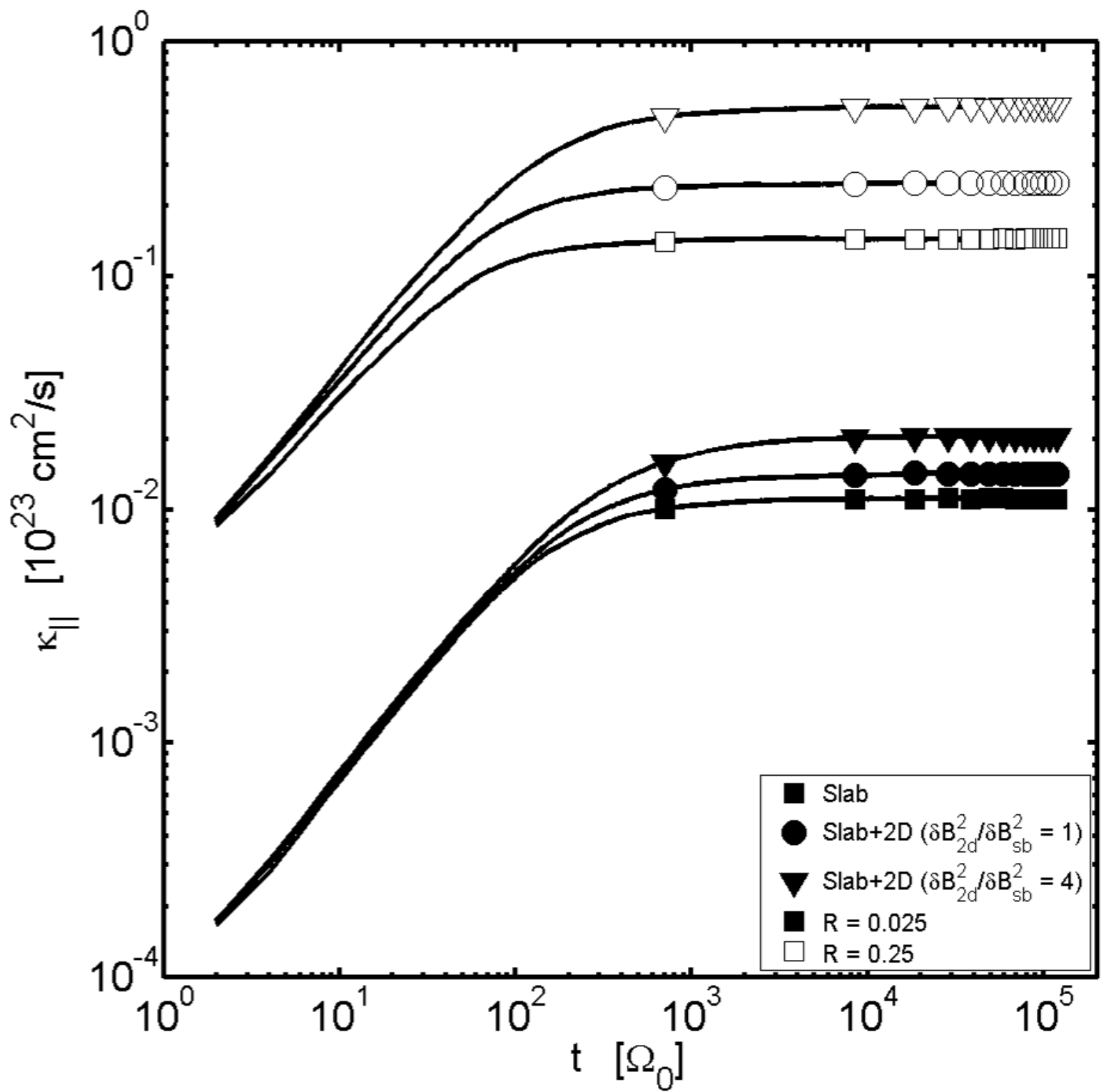

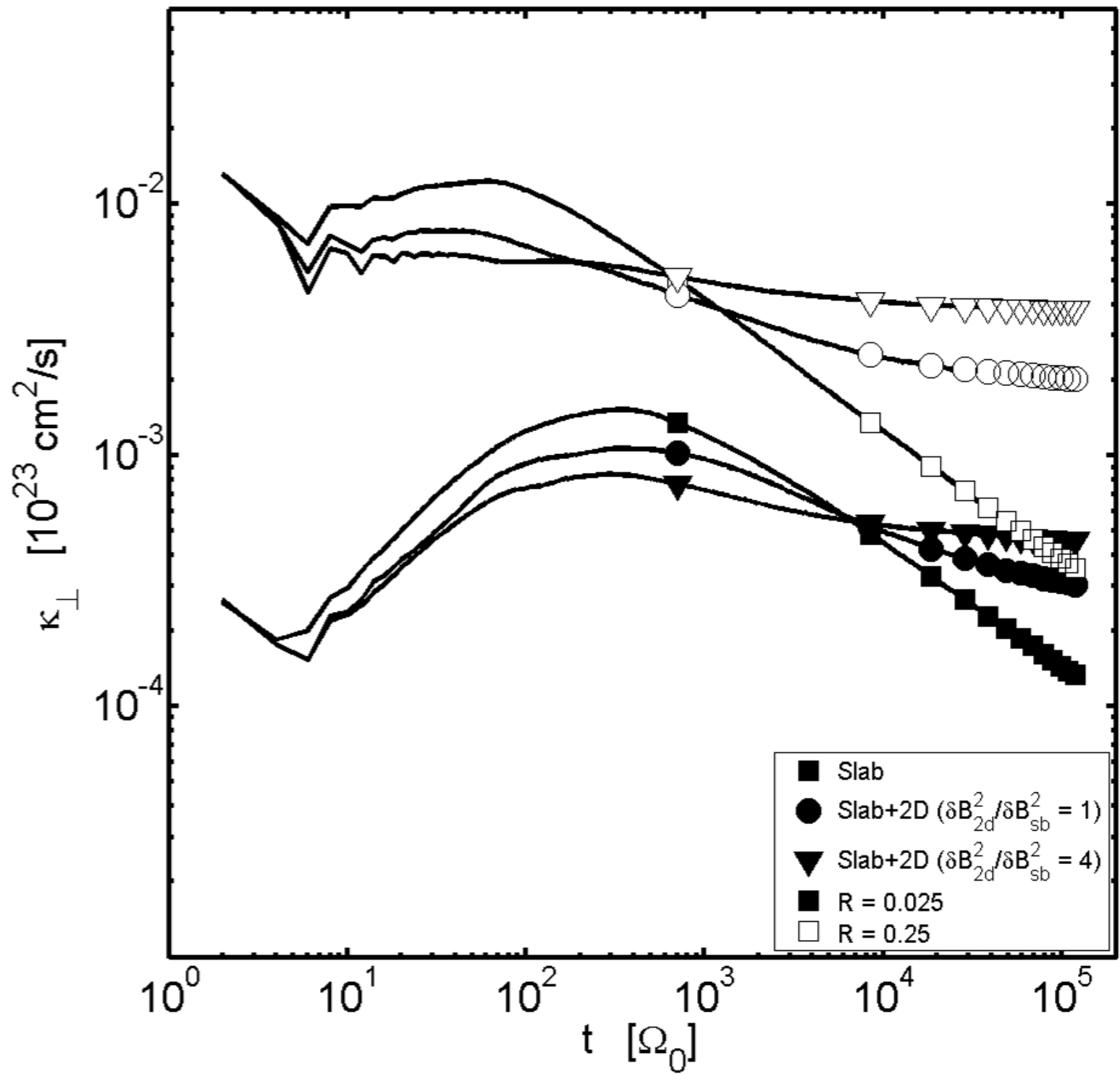

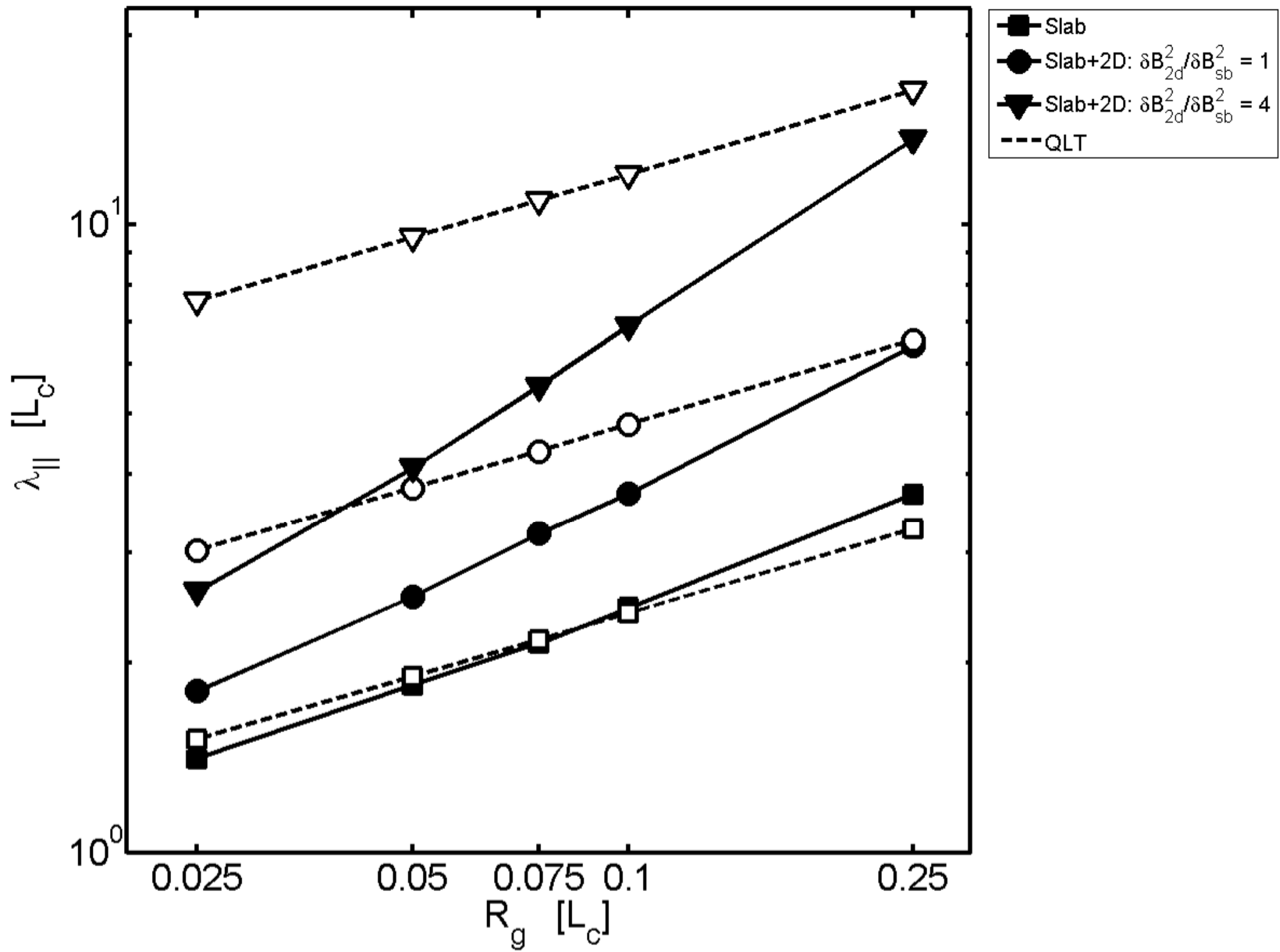

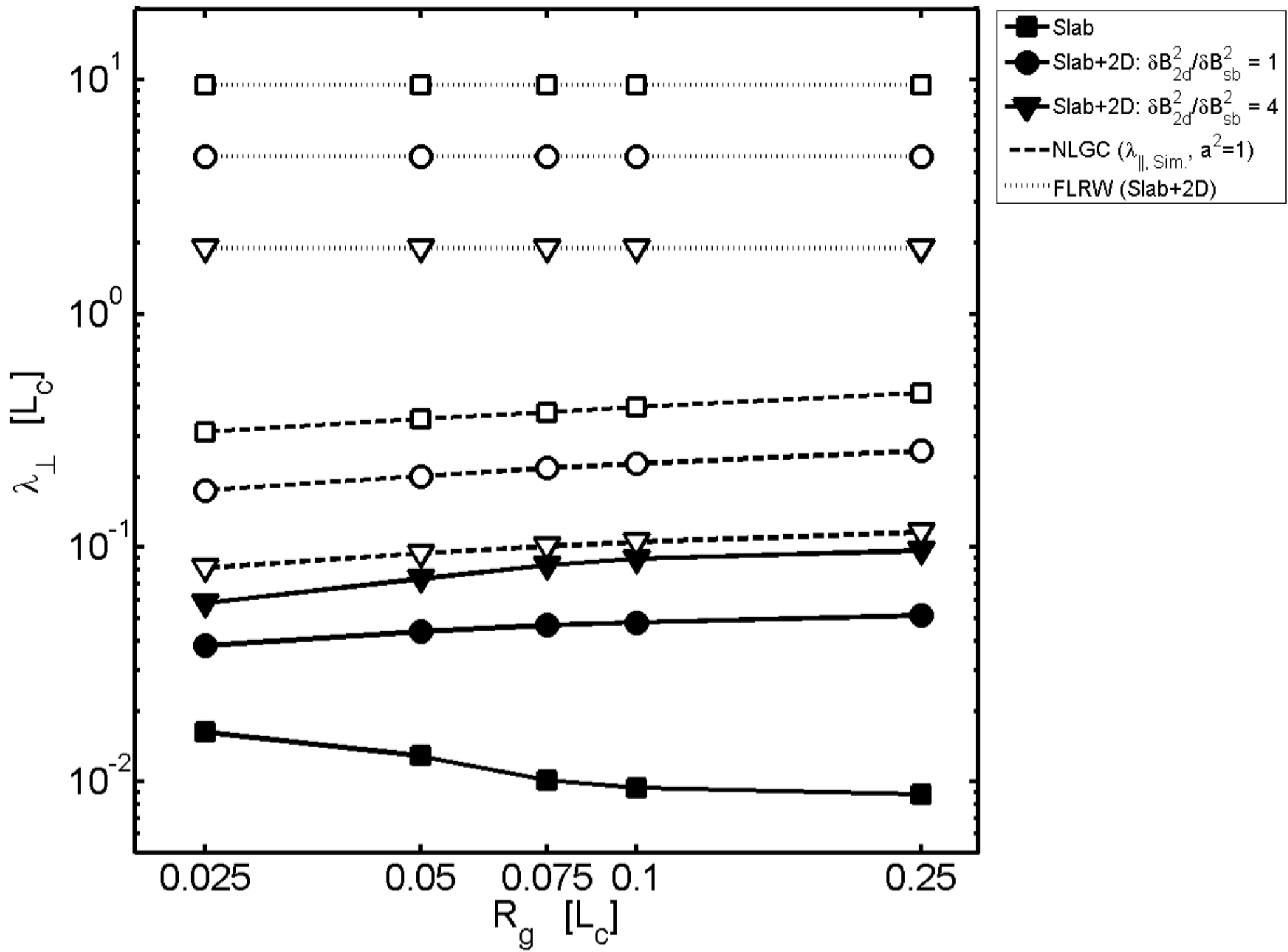

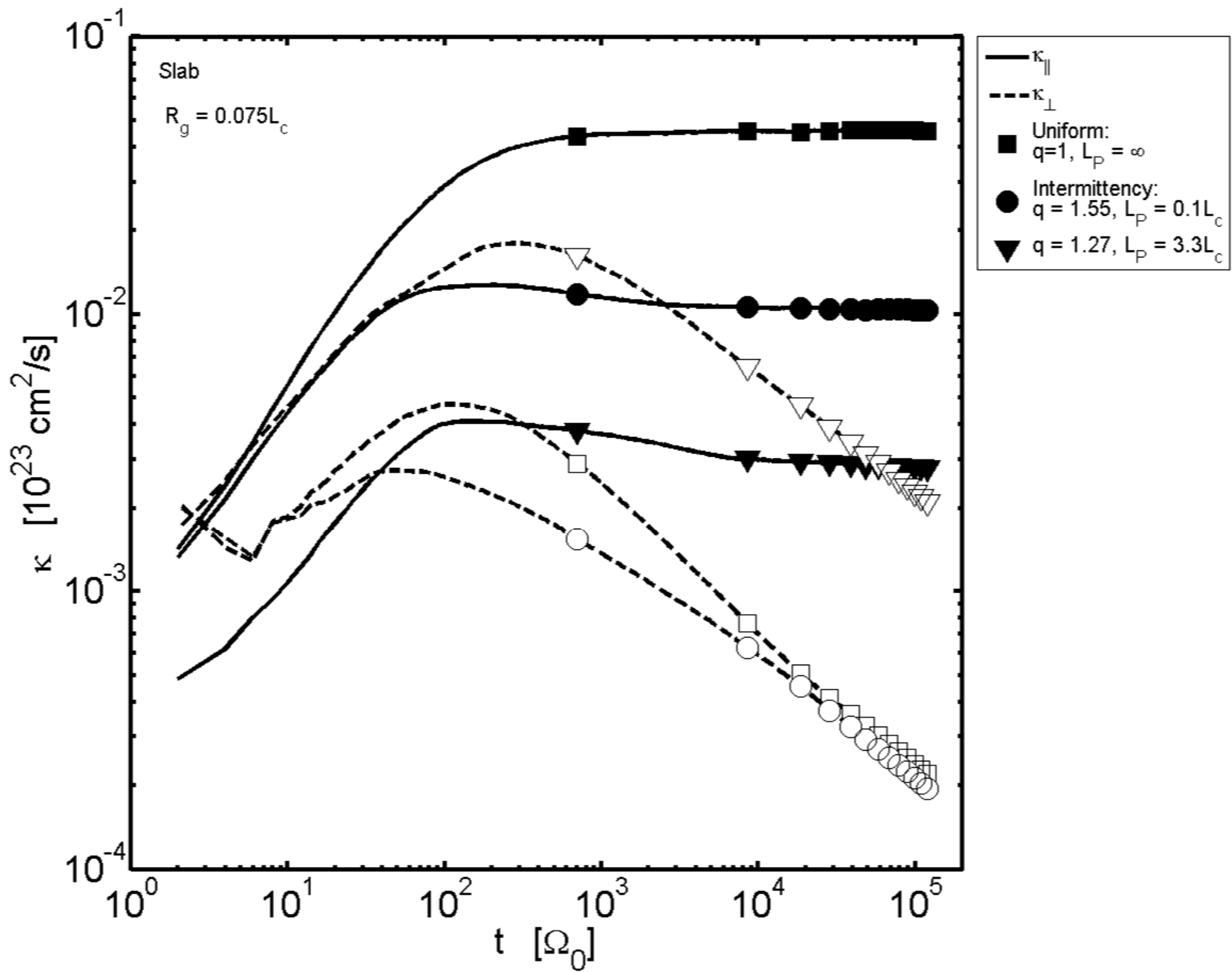

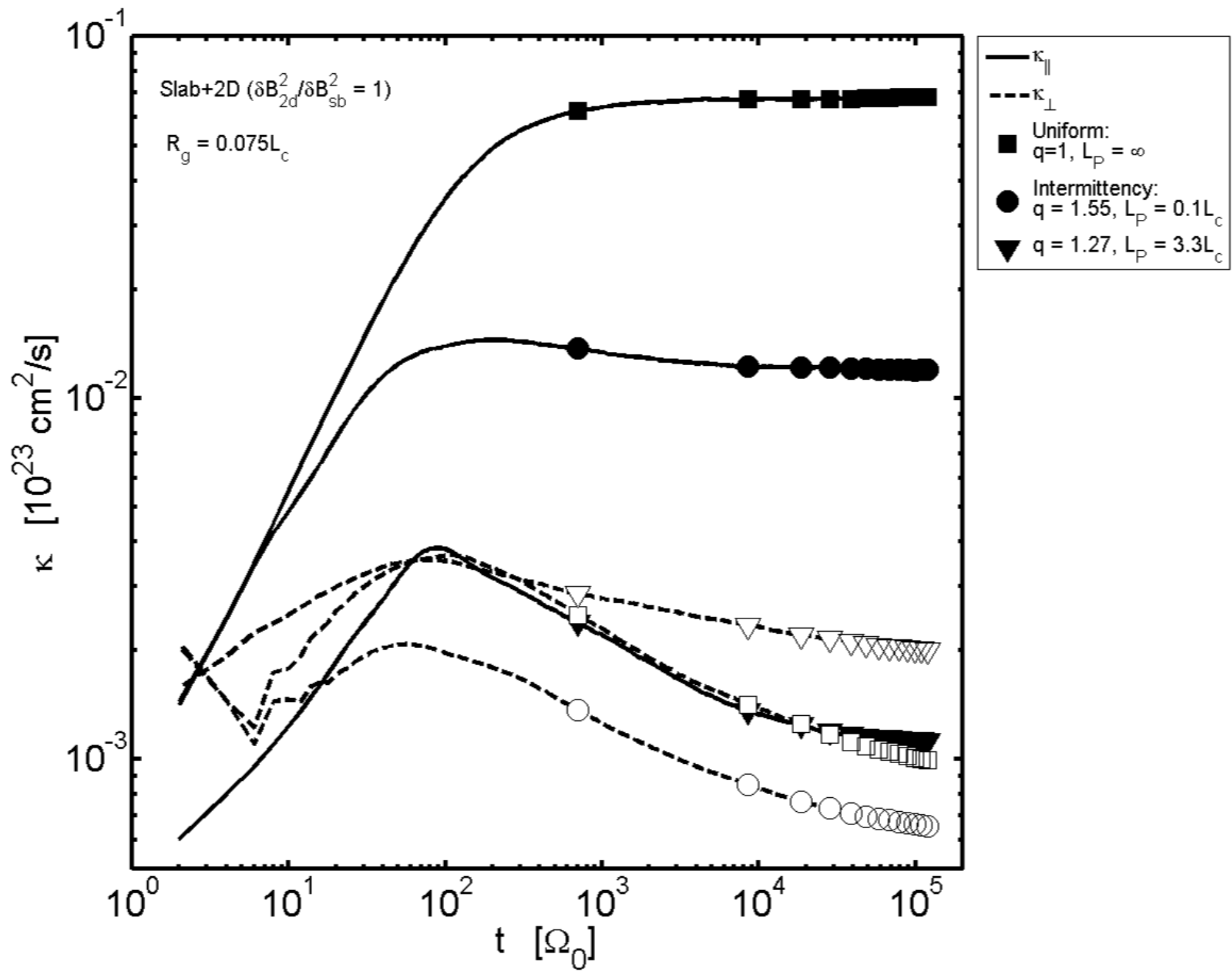

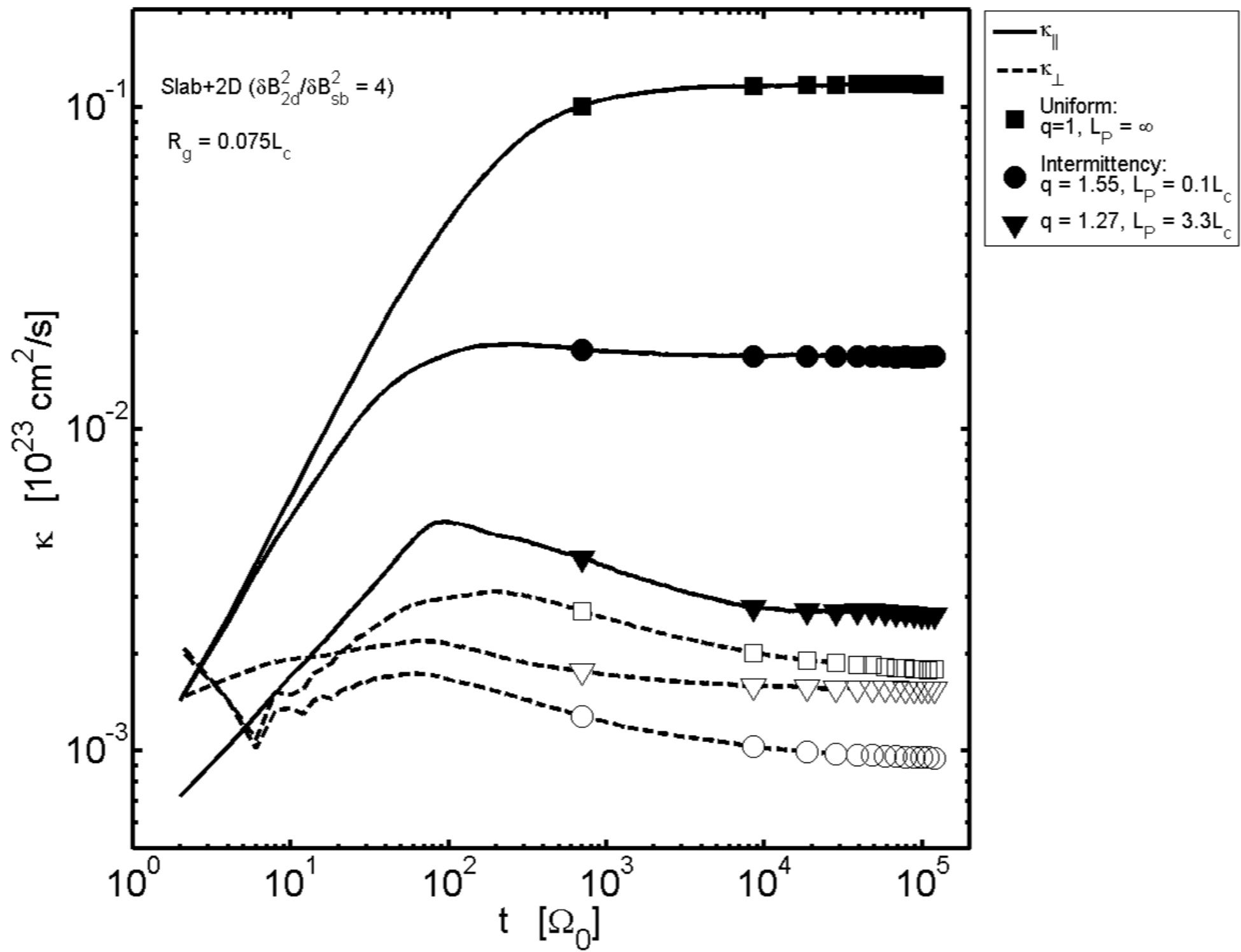

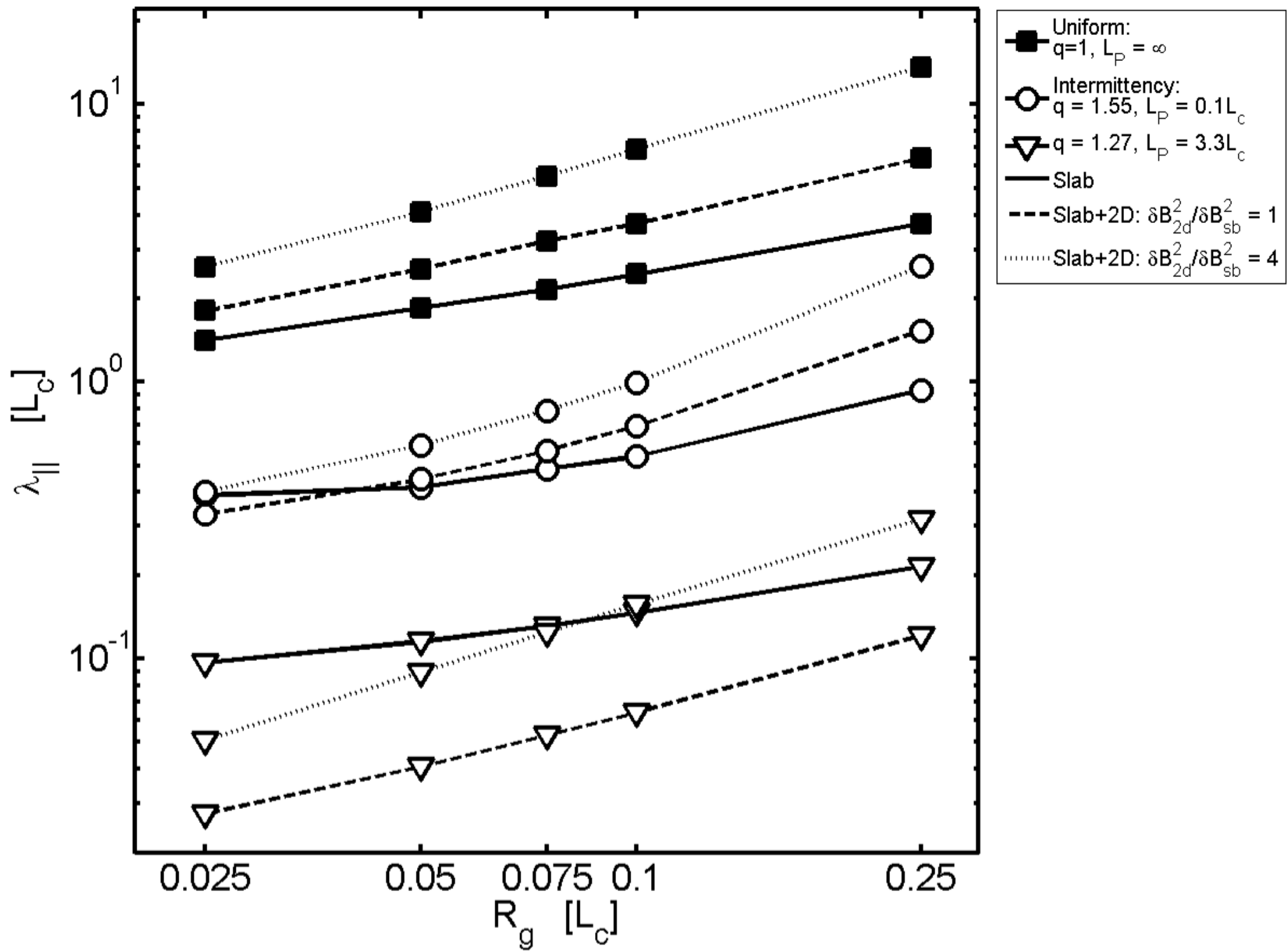

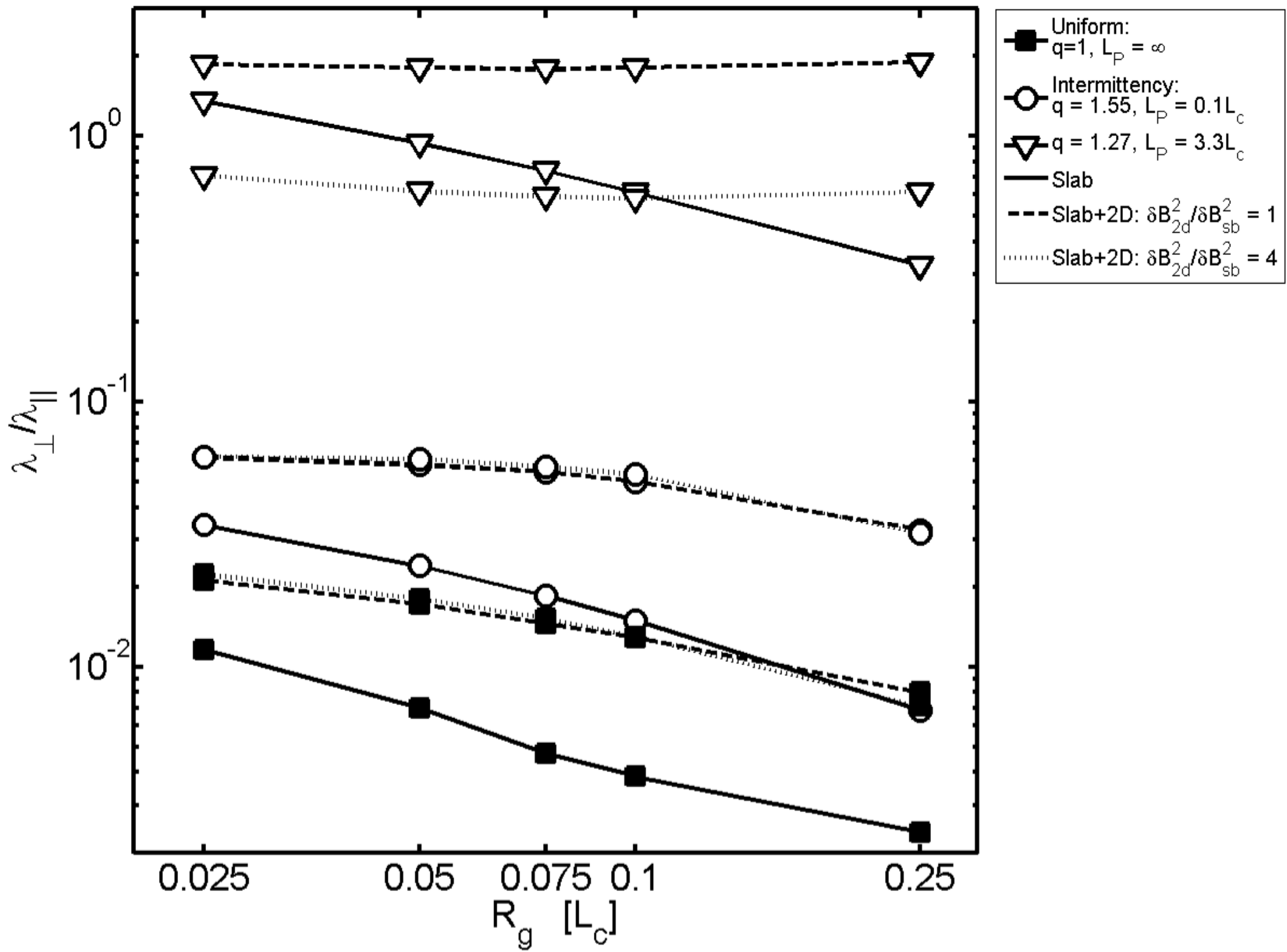

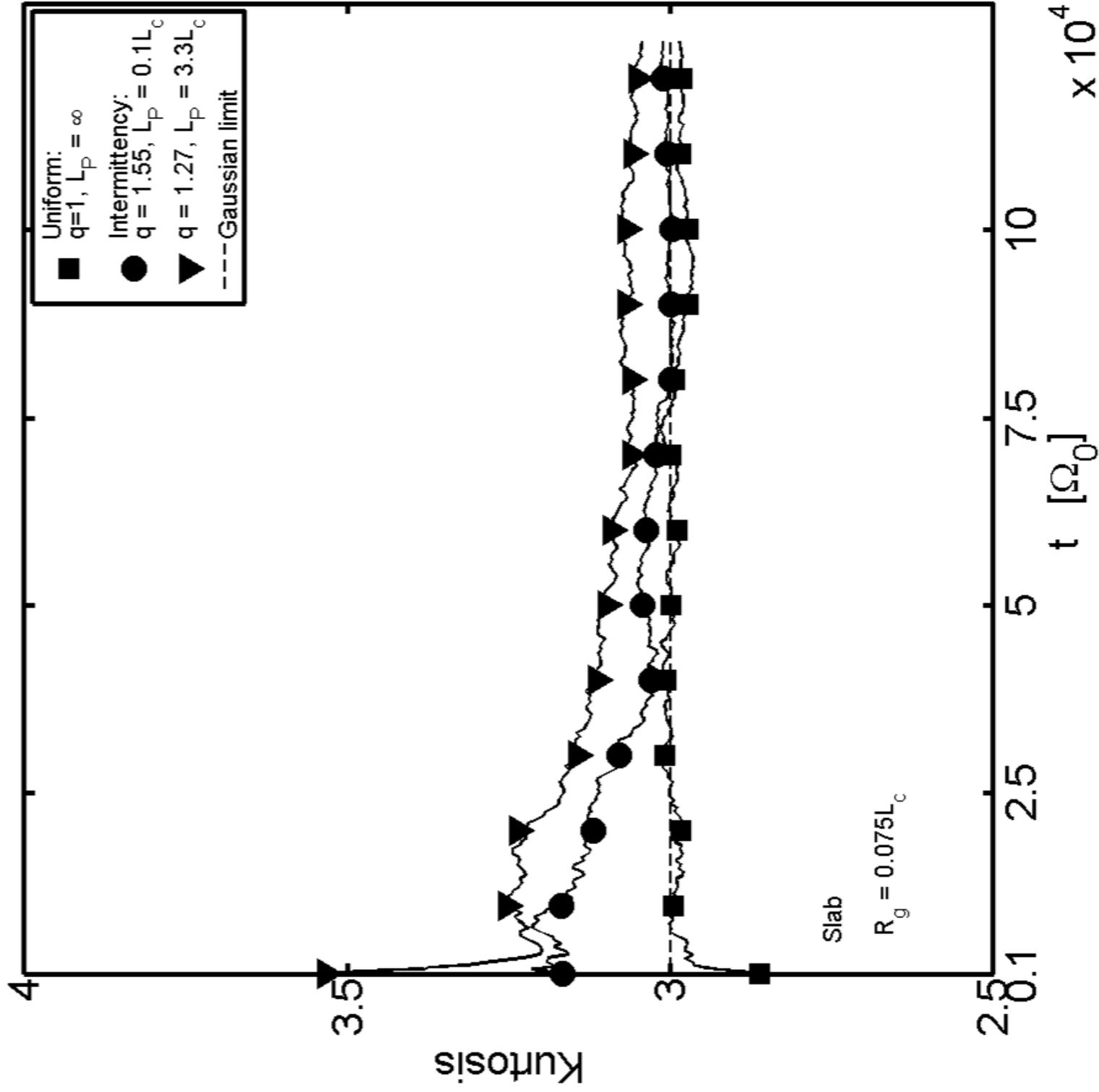

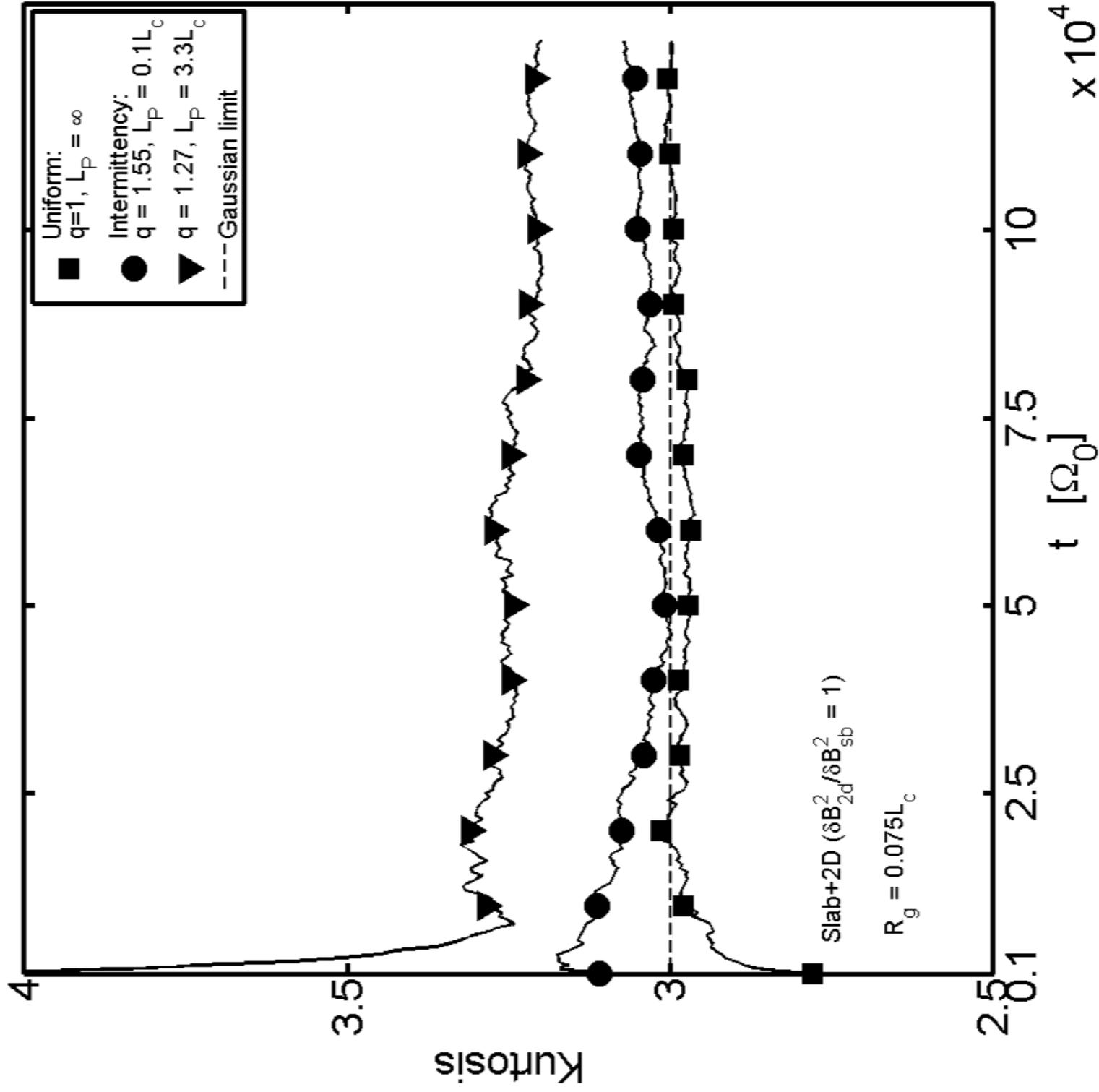

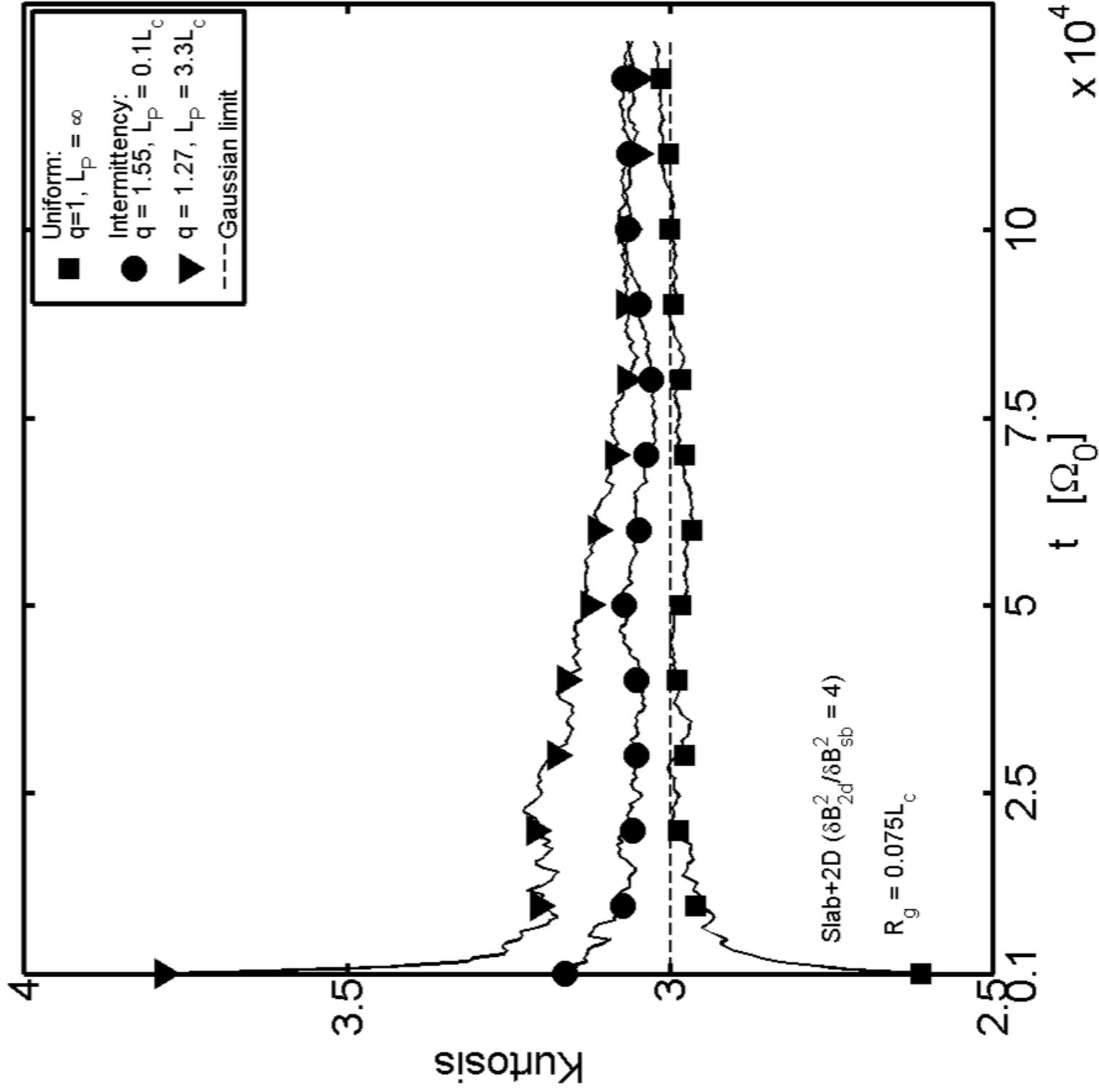